\documentclass[a4paper,11pt]{article}
\pdfoutput=1 

\usepackage{jheppub} 

\usepackage{amssymb} 
\usepackage{amsmath}
\usepackage{mathtools}
\usepackage{amsfonts}    
\usepackage{dsfont}
\usepackage{pdfpages}
\usepackage{verbatim}
\usepackage{tensor}
\usepackage{mathrsfs}
\usepackage[mathscr]{euscript}
\usepackage{tikz}
\usetikzlibrary{calc}
\usepackage{braket}
\usepackage{hyperref}

\usetikzlibrary{shapes.misc}
\tikzset{cross/.style={cross out, draw=black, ultra thick, minimum size=2*(#1-\pgflinewidth), inner sep=0pt, outer sep=0pt},
cross/.default={5pt}}

\usetikzlibrary{shapes.misc}
\tikzset{cross/.style={cross out, draw=black, ultra thick, minimum size=2*(#1-\pgflinewidth), inner sep=0pt, outer sep=0pt},
cross/.default={5pt}}

\let\oldbibliography\thebibliography
\renewcommand{\thebibliography}[1]{\oldbibliography{#1}
\setlength{\itemsep}{4.14pt}} 

\allowdisplaybreaks
\usepackage{mathrsfs}
\usepackage[mathscr]{euscript}

\usepackage{caption}
\usepackage{subcaption}

\numberwithin{equation}{section}

\def\be{\begin{equation}}
\def\ee{\end{equation}}

\title{\boldmath Twistor Coverings and Feynman Diagrams}

\author[a]{\!\! Faizan Bhat}
\author[b]{\!\!, Rajesh Gopakumar}
\author[b]{\!\!, Pronobesh Maity}
\author[c]{\!\! and Bharathkumar Radhakrishnan}

\affiliation[a]{Centre  for  High  Energy  Physics,  Indian  Institute  of  Science, \\ C.V.  Raman  Avenue,  Bangalore  560012,  India}
\affiliation[b]{International Centre for Theoretical Sciences-TIFR, \\
Shivakote, Hesaraghatta Hobli, Bengaluru North, 560 089, India}
\affiliation[c]{Département de Physique Théorique, Université de Genève, \\
24 quai Ernest-Ansermet, 1211 Genève 4, Suisse}

\emailAdd{faizanbhat@iisc.ac.in}
\emailAdd{rajesh.gopakumar@icts.res.in}
\emailAdd{pronobesh.maity@icts.res.in}
\emailAdd{bharathkumar.radhakrishnan@unige.ch}
\usepackage{tikz-cd}
\abstract{Recently, a worldsheet dual to free ${\cal N}=4$ Super Yang-Mills has been proposed in terms of twistor variables for ${\rm AdS}_5$, in parallel to that for the ${\rm AdS}_3$ dual to the free symmetric orbifold CFT. In the latter case, holomorphic covering maps play a central role in determining correlators and are associated to Feynman diagrams. After recasting these maps in terms of the worldsheet twistor variables for ${\rm AdS}_3$, we generalise to ${\rm AdS}_5$. We propose stringy incidence relations and appropriate reality conditions for the twistor covering maps. For some special kinematic configurations of correlators, we exhibit an explicit construction of the corresponding covering map. We find that the closed string worldsheet corresponding to this map is related to a gauge theory Feynman diagram by the Strebel construction, as for ${\rm AdS}_3/{\rm CFT}_2$. Rather strikingly, the regularised Strebel area of the worldsheet reproduces the Feynman propagator of the free field theory.}

\usepackage{tikz}
\usetikzlibrary{shapes.geometric, arrows}

\tikzstyle{gray} = [rectangle, rounded corners, minimum width=3cm, minimum height=1cm,text width= 5 cm, text centered, draw=black, fill=lightgray!30] 

\tikzstyle{gray2} = [rectangle, rounded corners, minimum width=3cm, minimum height=1cm,text width= 6 cm, text centered, draw=black, fill=lightgray!30] 

\tikzstyle{teal} = [rectangle, rounded corners, minimum width=3cm, minimum height=1cm,text width= 6 cm, text centered, draw=black, fill=lightgray!30! teal!30!] 

\tikzstyle{green2} = [rectangle, rounded corners, minimum width=3cm, minimum height=1cm,text width= 6 cm, text centered, draw=black, fill=lightgray!30! green!30!] 

\tikzstyle{magenta} = [rectangle, rounded corners, minimum width=3cm, minimum height=1cm,text width= 4 cm, text centered, draw=black, fill=lightgray!30! magenta!30!] 

\tikzstyle{Rightarrow} = [thick,->,>=stealth]

\tikzstyle{Leftarrow} = [thick,<-,>=stealth]

\tikzstyle{Leftrightarrow} = [thick,<->, >=stealth]

\begin{document}
\maketitle
\flushbottom

\section{Introduction}

Understanding the string dual to free ${\cal N}=4$ super Yang-Mills theory would give a new vantage point from which one could set out to decipher the AdS/CFT correspondence; this being the diametrically opposite regime from that described by supergravity on the dual ${\rm AdS}_5\times S^5$ spacetime. Recently, in \cite{Gaberdiel:2021qbb, Gaberdiel:2021jrv}, a proposal for a worldsheet description of the dual tensionless string theory was made\footnote{A somewhat different topological string approach to free super Yang-Mills is  proposed in \cite{Berkovits:2007zk, Berkovits:2007rj, Berkovits:2008qc, Berkovits:2019ulm}.}. This builds on the success of the description of the corresponding tensionless limit for strings on ${\rm AdS}_3\times{\rm S}^3\times {\mathbb T}^4$ dual to the free symmetric product orbifold CFT, ${\rm Sym}^N({\mathbb T}^4)$\cite{Eberhardt:2018ouy, Eberhardt:2019ywk, Eberhardt:2020akk, Dei:2020zui, Gaberdiel:2020ycd, Knighton:2020kuh, Gaberdiel:2021kkp}\footnote{The tensionless limit for ${\rm AdS}_3$ i.e. with massless higher spin fields in the spectrum, occurs for $k=1$ units of NS-NS flux - see \cite{Gaberdiel:2017oqg, Ferreira:2017pgt, Gaberdiel:2018rqv, Giribet:2018ada} for some of the precursor works at this special point. For works exploring the $k<1$ region, the transition to the $k>1$ region and their CFT duals, see \cite{Giveon:2005mi, Balthazar:2021xeh, Martinec:2021vpk, Eberhardt:2021vsx}.}. Both tensionless string descriptions are in terms of a set of free twistor variables, subject to a gauge constraint, with spectrally flowed sectors in the Hilbert space. These spectrally flowed sectors are crucial in accounting for the rich spectrum of single trace operators of the dual large $N$ CFT. While a first principles quantisation of the worldsheet theories is yet to be carried out in both cases, one can argue, based on a few plausible assumptions on the physical state conditions, that the spectrum nontrivially matches on both sides \cite{Eberhardt:2018ouy, Gaberdiel:2021qbb, Gaberdiel:2021jrv}. 

Additional support, in the free super Yang-Mills example, comes from the fact that the proposal gives a covariant version of the BMN \cite{Berenstein:2002jq} organisation of the large $N$ spectrum. Furthermore, this proposal is a closed string cousin of the ambitwistor open string description \cite{Berkovits:2004hg, Mason:2013sva} of tree level ${\cal N}=4$ Yang-Mills gluon amplitudes \cite{Witten:2003nn}. The worldsheet realisation of higher spin symmetries in this proposal has also been studied in \cite{Ahn:2021uri}. 

In the ${\rm AdS}_3/CFT_2$ case, one can, in fact, go beyond the agreement of the spectrum and argue for the (manifest) equality of correlation functions on both sides, giving a {\it de facto} derivation of the correspondence \cite{Eberhardt:2019ywk, Eberhardt:2020akk, Dei:2020zui}. On the worldsheet this followed from the localisation, on the string moduli space, of physical correlators onto those discrete points which admit finite degree holomorphic covering maps from the worldsheet to the spacetime (more accurately, its ${\rm S}^2$ boundary) with specified branching data. This localisation was later seen to follow from a stringy incidence relation obeyed by the twistorial worldsheet fields \cite{Dei:2020zui, Knighton:2020kuh}. This striking property on the worldsheet is precisely what is needed to match the Lunin-Mathur computation \cite{Lunin:2001} of correlators in the free symmetric orbifold CFT \cite{Eberhardt:2019ywk}. The latter computation is in terms of the same covering maps from an auxiliary Riemann surface (now identified with the worldsheet, cf. \cite{Pakman:2009}) to the boundary ${\rm S}^2$.   

The tensionless ${\rm AdS}_3/CFT_2$ example also gives a rather explicit realisation \cite{Gaberdiel:2020ycd} of a general mechanism for implementing open-closed string duality \cite{Gopakumar:2003ns, Gopakumar:2004qb, Gopakumar:2005fx}. This mechanism proceeds via a recasting of individual Feynman diagrams of the field theory into worldsheets of the dual closed string theory using the Strebel parametrisation of moduli space \cite{Strebel}. The upshot of \cite{Gaberdiel:2020ycd} was that, at least for correlators with large twist, the localisation of the previous para, was precisely to the points on moduli space which admit an {\it integer} Strebel differential - see Fig. \ref{fig:flowchart} for a flowchart of the logic of this correspondence. Strebel differentials are special meromorphic quadratic differentials $\phi_S(z) dz^2$ which are completely characterised by real ``lengths" $l_{ij}$ (defined as $\int_{a_i}^{a_j} dz\sqrt{\phi_S(z})$ between zeroes $(a_i, a_j)$); the Strebel lengths give a real parametrisation of the string moduli space. For an integer Strebel differential these lengths are proportional to (positive) integers\footnote{These are very special points in the moduli space which correspond to arithmetic Riemann surfaces (see for instance \cite{mulpenk1}) and also arise in the worldsheet dual to the Gaussian matrix model \cite{Razamat:2008zr, Gopakumar:2011ev, Gopakumar:2012ny}.}.  

\begin{figure}[!htb]
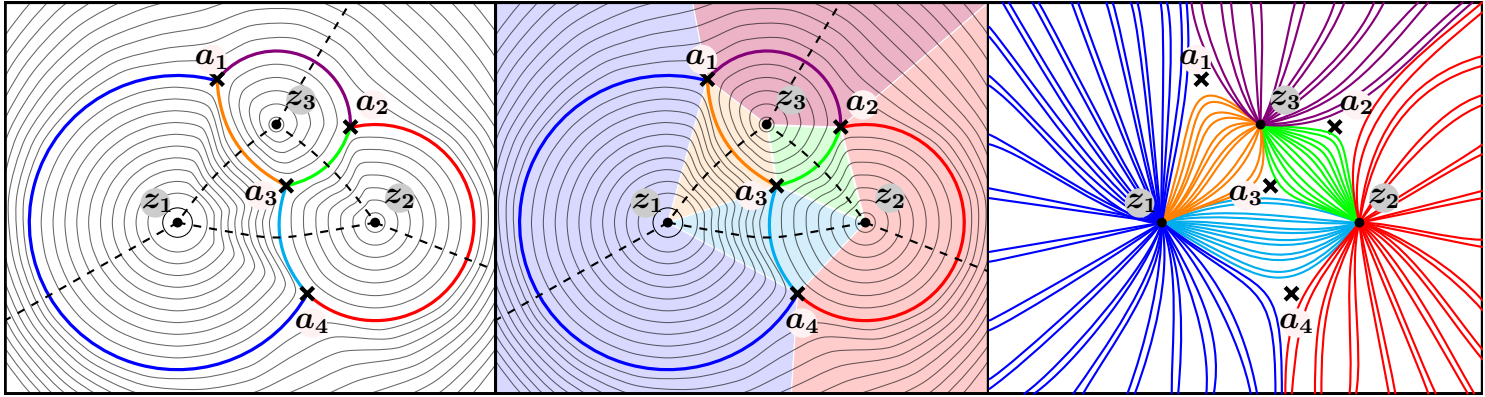

\hspace*{1.9em}
  \begin{subfigure}[b]{0.33\textwidth}
  \hspace*{-31mm}

 \end{subfigure}
    \caption{Open-Closed String Duality and the Strebel Construction. (Left) A foliation of a closed string Riemann Surface $\Sigma_{g,n}$ by the horizontal trajectories of the associated  Strebel differential. The coloured lines are `critical' horizontal trajectories with $a_k$ the zeroes of the differential. They form the Strebel graph. The $z_i$ are double poles and $z_4=\infty$ (not shown). (Middle) The Strebel graph gives a canonical decomposition (`gluing')  of $\Sigma_{g,n}$ into (shaded) regions which are conformal to infinite strips. The Strebel lengths are the widths of these strips. (Right) These strips are viewed as open string diagrams (Feynman-'tHooft double lines) which give rise to the free field Wick contractions. The width of the strips is identified with the number of contractions. Note that the skeleton Feynman graph (from gluing together homotopic edges), denoted by the dashed line in the middle and left, is the dual to the Strebel graph.}
    \label{fig:triptych}
\end{figure}

In the large twist limit of \cite{Gaberdiel:2020ycd}, this integer Strebel differential arises naturally as the Schwarzian of the covering map. As a consequence, one can identify the integer Strebel lengths directly with the number of ``Wick contractions" between vertices of the Feynman diagram that is associated to each covering map \cite{Pakman:2009} contribution to the field theory correlator. In other words, there is a one to one correspondence between the Feynman diagrams of the field theory and the individual closed string worldsheets that the string correlator localises onto. This gives a precise realisation of the proposal of \cite{Gopakumar:2005fx}, as refined by \cite{Razamat:2008zr}, where the Strebel differential is the bridge between the Feynman diagrams of the field theory and the worldsheets of the dual closed string. In fact, in this approach, the Strebel construction of the closed string worldsheet by gluing up strips with fixed Strebel lengths is the mathematical underpinning of open-closed string duality, with the strips being identified with the open string Feynman diagrams. See Fig. \ref{fig:triptych}. In the $AdS_3/CFT_2$ case, it was moreover seen that the closed string  weight associated to each Feynman diagram was the natural Nambu-Goto area for the worldsheet metric in Strebel gauge. 

In this paper, we will take steps towards generalising the above considerations and building a geometric picture in terms of twistor maps. We begin with $AdS_3$ and write down the classical twistor configurations which correspond to the holomorphic covering maps $\Gamma(z)$ that are the (exact) saddle points of the (genus zero) worldsheet path integral. These twistor covers are elegantly given in terms of the simple polynomials which enter into $\Gamma(z)$ (see Eqs.\eqref{twstrP}, \eqref{twstrQ}, \eqref{covmap}). This is consistent with the fact that only a finite number of wedge modes (for each twist/spectral flow label $w$) of the twistor fields are physical (Sec. 5 of \cite{Gaberdiel:2021jrv}) and therefore excited in the classical configuration\footnote{Recall that the wedge modes are those with mode number $r$ ($-\tfrac{w-1}{2} \leq r \leq \tfrac{w-1}{2}$). They behave as generalised zero modes in the $w$th spectrally flowed sector.}. We also find that the worldsheet stress tensor $T(z)$ of the twistor theory (Eq. \eqref{stress-tensor1}), evaluated on these twistor configurations is exactly the Schwarzian (and thus the Strebel differential, for large $w$) $S[\Gamma](z)$. This rounds off the picture for ${\rm AdS}_3$ in giving the twistor versions of the classical string configurations that the tensionless worldsheet path integral localises onto \cite{Eberhardt:2019ywk}. It will be interesting to see how the localisation onto these configurations arises directly from a worldsheet path integral for the twistor fields. 

We next proceed to consider the analogues of these solutions for ${\rm AdS}_5$. As per the proposal of \cite{Gaberdiel:2021qbb, Gaberdiel:2021jrv}, the worldsheet theory dual to the tensionless ${\rm AdS}_5\times {\rm S}^5$ now consists of the ambitwistor fields 
\begin{equation}\label{ambit0}
\begin{split}
    & Z^{I}=(\lambda^{\alpha},\mu^{\dot{\alpha}}; \psi^a)\\
    & Y_J=(\mu^{\dagger}_{\beta},\lambda^{\dagger}_{\dot{\beta}}; \psi_a^{\dagger})
    \end{split}
\end{equation}
subject to the ambitwistor constraint $Y_IZ^I =0$. We will drop, from now on, the fermionic fields which parametrise the ${\rm S}^5$ and focus on the bosonic twistor variables $(\lambda^{\alpha},\mu^{\dot{\alpha}})$ and $(\mu^{\dagger}_{\beta},\lambda^{\dagger}_{\dot{\beta}})$. 
These fields can be viewed, on the one hand, as parametrising the ambitwistor space of the conformal boundary of ${\rm AdS}_5$ or, on the other hand,  the twistor space of ${\rm AdS}_5$. We employ here the twistor description of (euclidean) ${\rm AdS}_5$ and its incidence relations following \cite{Adamo:2016rtr} (for our notation and conventions, see Appendix \ref{appA} and \cite{Adamo:2017qyl}). We are thereby led to propose a set of stringy incidence relations that the twistor worldsheet configurations should obey in the tensionless limit. 
\be\label{stringy-bulk0}
\begin{split}
        \mu^{\dagger}_{\alpha}(z)+X_{\alpha}^{\;\;\;\dot{\beta}}(z, \bar{z})\lambda^{\dagger}_{\dot{\beta}}(z) &= \lambda^{\beta}(z) \epsilon_{\beta \alpha}\\
        \mu^{\dot{\alpha}}(z)-X^{\dot{\alpha}}_{\;\;\; \beta}(z, \bar{z})  \lambda^{\beta}(z) &=\frac{1}{2}R^2(z, \bar{z}) \epsilon^{\dot{\alpha}\dot{\beta}} \lambda^{\dagger}_{\dot{\beta}}(z) \ .
    \end{split}
\ee
Here, $X^{\dot{\alpha}}_{\;\;\;\beta}(z, \bar{z})$ are the stringy coordinates along the $d=4$ boundary (in bispinor notation) while $R(z, \bar{z})$ is the radial profile. One may view these  stringy incidence relations as a general solution of the (bosonic) ambitwistor constraint (see Eq.\eqref{bulk-amb}) which must be imposed in the worldsheet theory as a gauge constraint. While the relations can be thus motivated, we expect them to be directly derived from the properties of worldsheet correlators as in \cite{Dei:2020zui}. This is currently under investigation \cite{GGKM}.
 
Imposing reality conditions on the $X^{\dot{\alpha}}_{\;\;\;\beta}(z, \bar{z})$ appropriate to the euclidean signature implies a natural set of reality conditions on the twistor fields. This is a direct indication that the (physical) left and right moving modes on the worldsheet are not independent as already suggested in \cite{Gaberdiel:2021qbb, Gaberdiel:2021jrv}. These reality conditions also enable us to invert the incidence relations and express the spacetime string configurations in terms of the twistor configurations. We will often restrict to configurations where the worldsheet is essentially at the boundary of the ${\rm AdS}_5$, as was the case for ${\rm AdS}_3$. However, unlike for ${\rm AdS}_3$, the boundary spacetime string configurations are in general, non-holomorphic. Nevertheless, there is a hidden holomorphy inherited from that of the underlying twistor fields. Thus, we immediately observe from the structure of Eq.\eqref{stringy-bulk0} that the matrix $\bar{\partial}X_{\alpha}^{\;\;\;\dot{\beta}}(z, \bar{z})$ has one zero eigenvalue.   

We then specialise to a special kinematic regime where the dual field theory operators are inserted at points $x_i$ all of which lie in a two dimensional plane (note that this is not a restriction for 2- and 3-point correlators). In this case, the holomorphic eigenvector picks out (together with the radial direction) an $AdS_3$ subspace. We can write down the twistor configuration again in terms of polynomials, whose ratio gives a {\it bona fide} holomorphic covering map. The polynomial nature of the configuration is again a reflection of the fact that only the wedge modes of the twistor fields are physical, as proposed in \cite{Gaberdiel:2021qbb, Gaberdiel:2021jrv}.  

The simplest such polynomial corresponds to the two point function of a BPS highest weight state in the SYM theory. In this case, from the explicit form of the covering map, we see that its Schwarzian, in the natural coordinates, is a constant which agrees with the Strebel differential in this case. The Strebel  length is proportional to the number of Wick contractions. Further, we compute the Strebel area  i.e. the area computed with the worldsheet metric given by $ds^2=|\phi_S(z)||dz|^2$ -- ``Strebel gauge" \cite{Gopakumar:2011ev}. This is formally infinite but needs to be evaluated with a careful regularisation. When we do that we find that a Nambu-Goto weight with the Strebel area precisely reproduces the free Feynman propagator associated to the Feynman diagrams! This picture can be extended to a multipoint correlator by gluing together the Strebel differentials in different patches, which we then identify with the Schwarzian of the covering map. The {\it additive} nature of the Strebel areas for each strip directly leads to the {\it multiplicative} form of the individual Feynman propagators. 

We view this as evidence that, in parallel to the ${\rm AdS}_3/CFT_2$ example, there exists a closed string picture in terms of twistor covering maps underlying Feynman diagrams in the free field limit. In particular, the fact that twistor covers of ${\rm AdS}_5$ reproduce the Feynman propagator, through the Strebel construction, buttresses the twistor proposal \cite{Gaberdiel:2021qbb, Gaberdiel:2021jrv} for the worldsheet description of free  ${\cal N}=4$ super Yang-Mills theory. Admittedly, we are working in a special kinematic regime, for the 4-point and higher correlator, but we expect to be able to overcome this restriction in the future. 

The plan of this paper is as follows: Sec. \ref{sec.2} focusses on the ${\rm AdS}_3$ case, where we exhibit  the twistor configurations corresponding to a covering map. After a brief review of the twistor geometry for ${\rm AdS}_5$ in Sec. \ref{sec.3}, we describe the euclidean reality conditions that we impose and some of its consequences. Sec. \ref{sec.4} uses these ingredients to propose stringy incidence relations for the worldsheet twistors. We then specialise to the classical configurations that correspond to mapping worldsheets into an ${\rm AdS}_3$ subspace. We employ the explicit form of these maps in Sec. \ref{sec.5} to 
flesh out the connection to Feynman diagrams in the gauge theory. We describe the resulting picture, which agrees with the Strebel construction, moreover reproducing the free propagator from the regularised Strebel area. Sec. \ref{sec.6} has brief concluding remarks. Appendix \ref{appA} sets out our twistor conventions and the nature of the twistor correspondence, with or without the Euclidean reality conditions.  Appendix \ref{appB} gives a quick recap, following \cite{Gaberdiel:2020ycd}, of the Strebel construction for tensionless ${\rm AdS}_3/CFT_2$, as a ready reference to compare with the logic of Section \ref{sec.5}.

\section{Twistor covering maps in $AdS_3$}\label{sec.2}

As we briefly recap in appendix \ref{appB}, correlators in the free symmetric product $CFT_2$ are given in terms of covering maps from an auxiliary covering space to the spacetime $S^2$, with specified branching data. In \cite{Eberhardt:2019ywk} it was shown that this is precisely mirrored in the tensionless string theory on $AdS_3\times S^3\times {\mathbb T}^4$ (i.e. with $k=1$ unit of NS-NS flux). Indeed, as anticipated in \cite{Lunin:2001, Pakman:2009} the covering space is the worldsheet of the dual string theory.  Thus the corresponding correlators in the worldsheet theory have the remarkable property that they are delta function supported on those points on the worldsheet moduli space which admit covering maps consistent with the branching data. The fundamental origin of this  localisation was later seen to be a twistorial ward identity for correlators in the free field realisation of the $\mathfrak{psu}(1,1|2)_1$ worldsheet theory \cite{Dei:2020zui, Knighton:2020kuh}. 

In  \cite{Eberhardt:2019ywk} it was also shown that when there is an admissible covering map, there is a semiclassical solution for the string worldsheet which exhibits this covering of the boundary $S^2$ of the $AdS_3$ part of the spacetime. As will be reviewed below, since the worldsheet is embedded in $AdS_3$ it also has a nontrivial radial profile even though it is essentially at the boundary. This gives a geometric picture, even in this highly stringy regime, of the string worldsheet exhibiting a nontrivial wrapping of the spacetime boundary. Since the underlying worldsheet theory is free in this limit, as can be seen in terms of the original sigma model (see Eq.~(2.4) and below of \cite{Eberhardt:2019ywk}) or from the free field realisation of the $\mathfrak{psu}(1,1|2)_1$ theory, these solutions are semiclassically exact. In fact, one can show \cite{Eberhardt:2019ywk} that the semiclassical action associated to these solutions gives precisely the Lunin-Mathur weight associated to each covering map.  

In this section, we will translate the solutions of \cite{Eberhardt:2019ywk}, which were given in terms of the $AdS_3$ coordinates, into classical solutions for the twistor fields that describe the $\mathfrak{psu}(1,1|2)_1$ theory. We will see that the solutions have a nice form which is also consistent with the behaviour of quantum correlators of these fields as found in \cite{Dei:2020zui}. We will also show that the classical stress tensor evaluated on these solutions agrees with the Schwarzian derivative of the covering map. Thus the stress tensor of the free field realisation is closely related to the Strebel differential due to the connection between the latter and the Schwarzian (at least in the limit of large twist). These solutions will give some guidance when we generalise to the case of the twistor classical solutions for the worldsheet theory of tensionless strings in $AdS_5$ in the next section.  

The conventional $AdS_3$ sigma model action, in first order form, is given by 
\be\label{adsact2}
S_{\mathrm{AdS}_3}= \frac{k}{4\pi} \int \mathrm{d}^2 z\  (4 \partial \Phi \bar{\partial} \Phi  +\bar \beta\partial  \bar{\gamma} +\beta \bar{\partial}  \gamma - \mathrm{e}^{-2\Phi} \beta \bar\beta-k^{-1} R \Phi)  \ . 
\ee
Here $\gamma(z)$ (and its conjugate) represent holomorphic coordinates for the spacetime $S^2$'s with which we foliate euclidean $AdS_3$. The radial coordinate $\Phi(z, \bar{z})$ is related to the usual poincare coordinate as $r=e^{-\Phi}$, such that the boundary is at $r=0 \leftrightarrow \Phi=\infty$. 
In terms of these coordinates one finds classical solutions by considering the holomorphic $\mathfrak{sl}(2, \mathbb{R})$ currents, as given by the Wakimoto form and imposing the right boundary conditions at the insertions together with the condition on the spacetime energies for the vertex operators (the $J^3_0$ eigenvalue near each insertion).  
\begin{equation}\label{sl2curr}
    \begin{split}
    &    J^{+}=\beta\\
     & J^3=-\partial \Phi +k\beta \gamma \\
     & J^{-}=-2\gamma \partial \Phi +\beta \gamma \gamma -\partial \gamma \ .
    \end{split}
\end{equation}
Here we have put the level $k=1$ in the usual (classical) Wakimoto representation. 
Thus the classical solution for an  $n$-point correlator of vertex operators for the ground states of twisted sectors of the dual symmetric orbifold CFT, is \cite{Eberhardt:2019ywk} 
\be\label{classol}
\gamma(z) = \Gamma(z)  \ , \qquad 
\Phi(z,\bar{z})= -\log\epsilon - \frac{1}{2}\ln{|\partial\Gamma|^2} \ , \qquad 
\beta(z) = -\frac{(\partial\Phi)^2}{\partial \Gamma} \ .  
\ee
$\Gamma(z)$ is the covering map, from the worldsheet to the spacetime, with branch points at the points $z_i$ where the vertex operators (in spectrally flowed sectors $w_i$) are inserted, with the behaviour as $z\rightarrow z_i$,
\be
\Gamma(z) = x_i+ a_i^{\Gamma}(z-z_i)^{w_i} +\ldots , \qquad (i=1\ldots n) \ .
\ee
 The covering map is uniquely specified by the branching data $(z_i, w_i)$ and specifying three of the $x_i$'s. Equivalently, if the $n$ locations $x_i$ are specified, the covering map exists only for discrete choices of $(n-3)$ of the $z_i$ i.e. on a discrete set of points on the moduli space ${\cal M}_{0,n}$.  
The radial field $\Phi(z, \bar{z})$ has an infrared cutoff $\epsilon \rightarrow 0$ which indicates that the string is stuck at the boundary ($\Phi=\infty$). We also see that $\partial\Phi$ is independent of $\epsilon$ and exhibits a nontrivial profile which will play an important role. 
Finally, the last relation in Eq.~(\ref{sl2curr}) for $\beta(z)$ follows from the on-shell condition for the worldsheet stress tensor which implies (for these ground state correlators) that 
\be\label{onshell}
J^+(z)J^-(z)= (J^3(z))^2 \ .
\ee

We now note that the $\mathfrak{sl}(2, \mathbb{R})$ currents in 
Eq.~(\ref{sl2curr}) are bilinears of the free twistor fields of $AdS_3$  (see Eq.~(2.2) of \cite{Dei:2020zui}). 
Thus in terms of the pairs of symplectic bosons $\xi^{\pm}$ and $\eta^{\pm}$ 
of $\mathfrak{u}(1,1|2)_1$ we have 
\begin{equation}\label{twstr-curr}
    J^3(z) = -(\eta^{+} \xi^{-})(z) \ ,  \quad J^{\pm}(z)= (\eta^{\pm} \xi^{\pm})(z)
\end{equation}
where we additionally have used the ambitwistor constraint which generates the quotient  
$\mathfrak{psu}(1,1|2)_1$
\begin{equation}\label{ambtwstr1}
    \xi^{+}\eta^{-}=\xi^{-}\eta^{+} \ .
\end{equation}
We notice that the on-shell constraint Eq.~(\ref{onshell}) is automatically satisfied in terms of the twistor variables.

We also note that this free field representation has the gauge freedom in which we rescale 
\be
\xi^{\pm}(z) \rightarrow \lambda(z)\xi^{\pm}(z) \ , \qquad \eta^{\pm} (z) \rightarrow \lambda(z)^{-1} \eta^{\pm}(z) \ .
\ee 
Together with the ambitwistor constraint Eq.~(\ref{ambtwstr1}), this means one can choose a gauge in which $\xi^{\pm}(z)=-\eta^{\pm}(z)$.  

We can then solve for $\xi^+$ and $\eta^+$ using the expression for $J^+(z)$ in Eq.~(\ref{sl2curr}) and equating it to that in Eq.~(\ref{twstr-curr}), together with the classical solution for $\beta(z)$ given in 
Eq.~(\ref{classol}). Then the expression for $J^3(z)$  allows us to solve for $\xi^-(z)$ and $\eta^-(z)$. We thus find the classical twistor solutions (in our gauge) to be 
\begin{equation}\label{classol-twstr}
       \xi^{+}=-\eta^{+}=-\frac{\partial\Phi}{\sqrt{\partial\Gamma}}= \frac{1}{2}\frac{\partial^2\Gamma}{(\partial\Gamma)^{\frac{3}{2}}},\quad \xi^{-}=-\eta^{-}=\frac{\Gamma \partial\Phi+\partial\Gamma}{\sqrt{\partial\Gamma}} = - \frac{\Gamma}{2}\frac{\partial^2\Gamma}{(\partial\Gamma)^{\frac{3}{2}}} +\sqrt{\partial\Gamma} \ .
\end{equation}
Note that the classical solutions obey stringy twistor ``incidence relations" in terms of the boundary variables. 
\be\label{incid1}
\xi^-+\Gamma\xi^+ =\sqrt{\partial\Gamma} = \sqrt{\rho}, \qquad \eta^-+\Gamma\eta^+ =-\sqrt{\partial\Gamma} = -\sqrt{\rho}\ .
\ee
The right hand side is not zero but rather proportional to the radial profile. Thus we have defined a holomorphic radial profile $\rho(z)$, using Eq.{\eqref{classol}, 
\be
r^2(z,\bar{z}) = e^{-2\Phi(z,\bar{z})} \equiv \epsilon^2\rho(z)\bar{\rho}(\bar{z}) \ .
\ee 
The RHS of Eq.{\eqref{incid1} vanishes as 
$(z-z_i)^{\frac{w_i-1}{2}}$ as one approaches any of the inserted vertex operators, where the string worldsheet is pinned to the boundary.

Given a covering map 
\begin{equation}\label{covmap}
    \Gamma(z) = \frac{P_N(z)}{Q_N(z)}
\end{equation}  
where $P_N(z)$ and $Q_N(z)$ are degree $N$ polynomials, define the Wronskian 
\be\label{wronsk} 
W = P'_N(z)Q_N(z) - Q'_N(z)P_N(z) = C \prod_{i =1}^n (z- z_i)^{w_i-1} \ . 
\ee
In the second equality we have used that $\partial\Gamma(z) = \frac{W(z)}{Q_N^2(z)}$ and vanishes as $(z- z_i)^{w_i-1}$ near each of the $z_i$. Comparing the degrees of the polynomials in Eq.~(\ref{wronsk}), we see that this determines the Wronskian upto the overall constant $C$. 

Then the $AdS_3$ twistors Eq.~(\ref{classol-twstr}) can be expressed nicely as
\be
\begin{split}
\xi^{+}  &= - \eta^{+} = \frac{1}{2 W(z)^{\frac{3}{2}}} ( W'(z) Q_N(z) - 2 W(z) Q_N'(z)) \\
 \xi^{-} &= - \eta^{-} = -\frac{1}{2 W(z)^{\frac{3}{2}}} ( W'(z) P_N(z) - 2 W(z) P_N'(z))
\end{split}
\ee
We also see that  
\begin{equation}
W'(z) = W(z) \sum_{i=1}^n\frac{ (w_i -1)}{z-z_i} = W(z) \frac{\tilde{R}_{n-1}(z)}{ \prod_{i =1}^n (z- z_i)}
\end{equation}
where $\tilde{R}_{n-1}(z) = \sum_{i=1}^n (w_i -1)\prod_{j \neq i}^n(z - z_j)$ is a polynomial of degree $(n-1)$ determined solely by the covering map data $(z_i, w_i)$. Using this, we can rewrite
\be
\begin{split}
W'(z) Q_N(z) - 2 W(z) Q'_N(z) &= W(z) \frac{\tilde{R}_{n-1}(z) Q_N(z) - 2 \prod_{i=1}^n (z-z_i) Q'_N(z)}{\prod_{i=1}^n (z-z_i)}\\
&=  \frac{W(z) \tilde{Q}_{N+n-1}(z)}{\prod_{i=1}^n (z-z_i)}
\end{split}
\ee
where $\tilde{Q}_{N+n-1}(z)$ is a polynomial of degree $(N+n-1)$ that is completely determined by $Q_N(z)$ and the covering map data $(z_i, w_i)$. In fact, notice that
\be 
-\frac{1}{2} \frac{\tilde{Q}_{N+n-1}(z)}{\prod_{i =1}^n (z- z_i)^{\frac{(w_i+1)}{2}}} = \frac{d}{dz} \Big[ \frac{Q_N(z)}{\prod_{i =1}^n (z- z_i)^{\frac{(w_i-1)}{2}}} \Big] \ .
\ee
We can similarly define 
\be\label{Pdef}
\tilde{P}_{N+n-1}(z) =\tilde{R}_{n-1}(z) P_N(z) - 2 \prod_{i=1}^n (z-z_i) P'_N(z) \ . 
\ee
\\
Then we can express the twistor classical solutions in a nice symmetric form as 
\be\label{twstrQ}
\xi^{+} = - \eta^{+} = \frac{\tilde{Q}_{N+n-1}(z)}{2\prod_{i =1}^n (z- z_i)^{\frac{(w_i+1)}{2}}} = - \frac{d}{dz} \Big[ \frac{Q_N(z)}{\prod_{i =1}^n (z- z_i)^{\frac{(w_i-1)}{2}}} \Big] 
\ee
and similarly,
\be\label{twstrP}
\xi^{-} = - \eta^{-} = -\frac{\tilde{P}_{N+n-1}(z)}{2\prod_{i =1}^n (z- z_i)^{\frac{(w_i+1)}{2}}} = \frac{d}{dz} \Big[ \frac{P_N(z)}{\prod_{i =1}^n (z- z_i)^{\frac{(w_i-1)}{2}}} \Big] \ .
\ee

The takeaway from the final form of the twistor solutions in Eqs.~(\ref{twstrQ}, \ref{twstrP}) is that they are rational functions which are determined by the two polynomials $Q_N, P_N$ and the covering data. This fits in with our expectation for the quantum theory that only the wedge modes of the twistors are excited on-shell -- see the discussion around Eq.~(2.26) of \cite{Gaberdiel:2021jrv}. We see that these twistors have a singularity as $z\rightarrow z_i$ 
\be
\xi^{\pm}, \eta^{\pm} \sim \frac{1}{(z- z_i)^{\frac{(w_i+1)}{2}}} \ .
\ee
This is consistent with what we know about the OPE of these fields in the quantum correlators. In particular, we know that at the origin ($x=0$) the fields behave as in Eqs.~(2.36-39) of \cite{Dei:2020zui}. However, at a generic point, the OPE of these fields with the spectrally flowed vertex operators is given by Eq.~(2.42) of \cite{Dei:2020zui}. And in that notation $\xi^{\pm(x)}, \eta^{\pm (x)}$ behave as 
$(z- z_i)^{-\frac{(w_i+1)}{2}}$. See also Eq.~(4.12) of  \cite{Dei:2020zui}. The combination which appears in the incidence relation then has a regular OPE which is also reflected in the vanishing of the right hand side of Eq.~(\ref{incid1}) as $z\rightarrow z_i$.

For the generalisation and comparison to the higher dimensional case we define the ambitwistor variables 
\begin{equation}
    Z^I=\begin{pmatrix}
        \xi^{+}\\ \xi^{-}
    \end{pmatrix},\quad 
    Y_I=\begin{pmatrix}
        -\eta^{-}\\ \eta^{+}
    \end{pmatrix}
\end{equation}
and the ambitwistor constraint Eq.~(\ref{ambtwstr1}) reads as  
\begin{equation}\label{ambitwistor_condition}
 Y_{I}Z^{I}=\xi^{-}\eta^{+}-\xi^{+}\eta^{-}=0 \ .   
\end{equation}

We also note for later reference that the stress tensor of the worldsheet theory is given by
\begin{equation}\label{stress-tensor1}
    \begin{split}
        T(z)&=\frac{1}{2}\epsilon_{\alpha \beta} (\xi^{\alpha}\partial \eta^{\beta}+\eta^{\alpha}\partial \xi^{\beta})\\
        &= -\eta^{-}\partial \xi^{+}+\eta^{+}\partial \xi^{-} = Y_I\partial Z^I
    \end{split}
\end{equation}
where we have used Eq.~(\ref{ambtwstr1}) in the second line. 
Evaluating this in terms of the classical solution Eq.~(\ref{classol-twstr}) leads to
    \begin{equation}
        \begin{split}
        T(z)&=\frac{1}{2}\left[ \partial\left(\frac{\partial^2\Gamma}{\partial\Gamma}\right)-\frac{1}{2} \left(\frac{\partial^2\Gamma}{\partial\Gamma}\right)^2 \right] \\ &=\frac{1}{2} S\left[ \Gamma(z)\right]
        \end{split}
        \end{equation}

In \cite{Gaberdiel:2020ycd}, it was shown that, in the large twist limit, the quadratic differential defined by the Schwarzian of the covering map could be identified with the integer Strebel differential which defines the corresponding closed string worldsheet. We therefore see that there is a close relation between the worldsheet stress tensor and the Strebel diffferential - a relation which we think should be generic in open-closed string duality\footnote{See remark 5.3 on p. 101 of \cite{Eynard-lec} for a reason why the two might be related in a `heavy' limit. We thank Edward Mazenc for drawing our attention to this.}.

\par

\section{The Twistor space of ${\rm AdS}_5$}\label{sec.3}

We want to generalise the twistor solutions of the previous section to the case of ${\rm AdS}_5$. 
As a preliminary, in this section, we will describe the twistor space of ${\rm AdS}_5$ and its relation to the ambitwistor space of its conformal boundary, following the discussion in \cite{Adamo:2016rtr}. We then review the twistor incidence relations in ${\rm AdS}_5$ as well as the reality condition that can be imposed in Euclidean signature. We will find that imposing these reality conditions enables one to explicitly solve for the usual spacetime coordinates of ${\rm AdS}_5$ in terms of the twistors. 

We begin with the complex projective space $\mathds{CP}^5$ with homogeneous co-ordinates $X^{IJ}$ represented as an antisymmetric $4\times4$ matrix with the identification $X\sim \lambda X$ with $\lambda\in \mathds{C}^{*}$. 
We can then define a holomorphic metric on $\mathds{CP}^5$ which has the property of being invariant under local scalings $X_{IJ}\to \lambda(X^2) X_{IJ}$, 
\begin{equation}\label{metric_CP^5}
    ds^2=-\frac{dX^2}{X^2}+\left( \frac{X\cdot d\,X}{X^2}\right)^2
\end{equation}
where the contraction of indices is performed w.r.t $\epsilon_{IJKL}$. 
This is the metric on complexified ${\rm AdS}_5$.

The conformal boundary of this spacetime is at $X^2=0$ i.e. 
\begin{equation}
    M=\{ X^2=0\,| \,X\in \mathds{CP}^5\} \ .
\end{equation}
Using the scaling freedom we can always parametrise the points in $M$ as, \cite{Adamo:2016rtr}
\footnote{For our conventions, especially with regard to spinor indices, we refer the reader to Appendix \ref{appA}.}
\begin{equation}
    X_b^{IJ}=\begin{bmatrix}
        \epsilon^{\alpha \beta}&& x^{\alpha \dot{\beta}}\\
        -x^{\dot{\alpha}\beta} && \frac{1}{2}x^2 \epsilon^{\dot{\alpha}\dot{\beta}} 
    \end{bmatrix} \ .
\end{equation}
This form ensures $X_b^2=0$ i.e $\det (X_b)=0$. It immediately gives
\be 
(X_b)_{IJ}=\begin{bmatrix}
        \frac{1}{2}x^2\epsilon_{\alpha\beta} && -x_{\alpha\dot{\beta}}\\
        x_{\dot{\alpha}\beta}&& \epsilon_{\dot{\alpha}\dot{\beta}} 
    \end{bmatrix} \ .
\ee

Going away from the boundary, in terms of,
\begin{equation}
    I^{IJ}= \begin{bmatrix}
           0 & 0\\
           0 & \epsilon^{\dot{\alpha}\dot{\beta}} 
    \end{bmatrix} \ ,
\end{equation}
we can parametrise a generic point as 
\begin{equation}\label{bulk-pt}
    X^{IJ}= (X_b)^{IJ}+\frac{r^2}{2}I^{IJ} = 
    \begin{bmatrix}
        \epsilon^{\alpha \beta}&& x^{\alpha \dot{\beta}}\\
        -x^{\dot{\alpha}\beta} && \frac{1}{2}(x^2+r^2) \epsilon^{\dot{\alpha}\dot{\beta}}
    \end{bmatrix} \ .
\end{equation}
We then have 
\begin{equation}\label{rad}
    X_{IJ}X^{IJ}=2r^2 \ .
\end{equation}
It is not difficult to verify that this parametrisation, when plugged into Eq. \eqref{metric_CP^5}, gives the usual Poincare metric, with $r^2$ being the radial coordinate.  We refer the reader to \cite{Adamo:2016rtr} for further details. 

We now define the twistors
\begin{equation}\label{ambit}
\begin{split}
    & Z^{I}=(\lambda^{\alpha},\mu^{\dot{\alpha}})=(\lambda^1,\lambda^2,\mu^1,\mu^2)\\
    & Y_J=(\mu^{\dagger}_{\beta},\lambda^{\dagger}_{\dot{\beta}})=(\mu^{\dagger}_1,\mu^{\dagger}_2,\lambda^{\dagger}_1,\lambda^{\dagger}_2)
    \end{split}
\end{equation}
These will play the role of ambitwistor variables for the conformal boundary of ${\rm AdS}_5$ but will more generally be viewed as twistor variables for the bulk. We next describe the incidence relations that they obey which define the twistor correspondence with  spacetime on the boundary as well as the bulk. 

\subsection{The incidence relation on the boundary} 

On the boundary the above twistor variables $(Z^I, Y_J)$ can be defined as being in the kernel of 
$(X_b)_{IJ}$ and its dual $X_b^{IJ}$, respectively \cite{Adamo:2016rtr}. Note that the kernel is non-empty precisely at the boundary where, as we noted $\det (X_b)=0$. 
Then we see that the relation $(X_b)_{IJ}Z_b^{J}=0$  yields
\be
\begin{split}
           x_{\dot{\alpha}\beta}\lambda^{\beta}+\epsilon_{\dot{\alpha}\dot{\beta}}\,\mu^{\dot{\beta}}&=0
         \\  \frac{1}{2}x^2\epsilon_{\alpha\beta} \lambda^{\beta}-x_{\alpha\dot{\beta}}\,\mu^{\dot{\beta}}&=0
    \end{split} \ .
\ee
In fact, the second equation is a consequence of the first, using the first identity in (\ref{identity}). The latter can also be expressed as
\be\label{boundary_incidence1}
\mu^{\dot{\alpha}}=x^{\dot{\alpha}}_{\;\;\;\beta} \lambda^{\beta} \ .
\ee
The dual kernel relation $X_b^{IJ}Y^b_J=0$ yields 
\begin{equation}
    \begin{split}
        \epsilon^{\alpha \beta} \mu^{\dagger}_{\beta}+x^{\alpha\dot{\beta}}\lambda^{\dagger}_{\dot{\beta}}&=0
        \\  -x^{\dot{\alpha}\beta}\mu^{\dagger}_{\beta}+\frac{1}{2}x^2 \epsilon^{\dot{\alpha}\dot{\beta}} \lambda^{\dagger}_{\dot{\beta}}&=0 \ .
    \end{split} 
\end{equation}
Again the second equation arises from the first and the latter can be neatly expressed as
\be\label{boundary_incidence2}
\mu^{\dagger}_{\alpha}=-x_{\alpha}^{\;\;\; \dot{\beta}} \lambda^{\dagger}_{\dot{\beta}} \ .
\ee
We also note that the incidence relation Eq.~\eqref{boundary_incidence1} and its dual Eq.~\eqref{boundary_incidence2} together imply that the $(Z_b^I, Y^b_J)$ obey the ambitwistor constraint 
\be\label{bulkC}
{\cal C}_b \equiv Z_b^I Y^b_I =0 \ ,
\ee
justifying the terminology. 

Using the incidence relations (\ref{boundary_incidence1}) and (\ref{boundary_incidence2}), we can represent a point on the boundary $M_{\mathds{C}}$ via any of the two bitwistors:
\begin{equation}\label{bitwstr}
\begin{split}
 (X_b)^{IJ}=-\frac{Z_1^{[I}Z_2^{J]}}{\langle \lambda_1 \lambda_2 \rangle}\quad \text{and}\quad
 (X_b)_{IJ}=-\frac{(Y_1)_{[I}(Y_2)_{J]}}{[\lambda^{\dagger}_1\lambda^{\dagger}_2]} 
\end{split}
\end{equation}
This is the usual twistor correspondence which associates a line in $\mathds{PT}$ (determined by two points $Z_1$ and $Z_2$) to a point in  complexified Minkowski space $M_{\mathds{C}}$. The second relation above is the analogous correspondence for the dual twistor space. We refer the reader to Figs. \ref{fig:fibration_on_boundary} and \ref{fig:fibration_in_bulk} in appendix \ref{appA} which describes the twistor correspondence as a double fibration, on both the boundary and in the bulk, in the complexified case as well as after imposing the (euclidean) reality conditions of Sec. \ref{sec.3.3}.

\par

\subsection{The incidence relation in the bulk}\label{sec.3.2}

The above incidence relations in the boundary arise from a careful limit of incidence relations in the bulk. When $r\neq 0$, we no longer have a nontrivial kernel for $X_{IJ}$. Instead we impose the natural twistorial incidence relation \cite{Adamo:2016rtr}
\begin{equation}\label{bulk_incidence1}
\begin{split}
Z^{I}&=X^{IJ}Y_J \\ \Rightarrow 
 \begin{bmatrix}
        \lambda^{\alpha}\\ \mu^{\dot{\alpha}}
 \end{bmatrix}&=\begin{bmatrix}
        \epsilon^{\alpha \beta}&& x^{\alpha \dot{\beta}}\\
        -x^{\dot{\alpha}\beta} && \frac{1}{2}(x^2+r^2) \epsilon^{\dot{\alpha}\dot{\beta}}
    \end{bmatrix} \begin{bmatrix}
           \mu^{\dagger}_{\beta} \\
           \lambda^{\dagger}_{\dot{\beta}}
    \end{bmatrix} \ .
    \end{split}
\end{equation}
This gives the following two independent equations:
\begin{equation}\label{bulk_incidence2}
    \begin{split}
        \mu^{\dagger}_{\alpha}+x_{\alpha}^{\;\;\;\dot{\beta}}\lambda^{\dagger}_{\dot{\beta}} &= \lambda^{\beta} \epsilon_{\beta \alpha}\\
        \mu^{\dot{\alpha}}-x^{\dot{\alpha}}_{\;\;\; \beta} \lambda^{\beta} &=\frac{1}{2}r^2 \epsilon^{\dot{\alpha}\dot{\beta}} \lambda^{\dagger}_{\dot{\beta}}
    \end{split}
\end{equation}
where the second relation is obtained using the second identity in (\ref{identity}). 
We also note that the bulk incidence relation in the form $Z^{I}=X^{IJ}Y_J$ immediately implies the quadric (or ambitwistor) relation 
\be\label{bulk-amb}
{\cal C} = Z^IY_I=0
\ee
 since the $X^{IJ}$ is antisymmetric in its indices. Thus the incidence relations can be viewed as disentangling the ambitwistor constraint. 

To recover the boundary incidence relations in Eqs.~\eqref{boundary_incidence1} and \eqref{boundary_incidence2} as $r\to 0$ we need to take the limit where we scale 
$Y_J \to \frac{Y^b_J}{r}$ whereas $Z^I\to Z_b^I$. In other words, $(\mu^{\dagger}_{\beta}, \lambda^{\dagger}_{\dot{\beta}})$ scale as $\frac{1}{r}$ as we approach the boundary, while 
$(\mu^{\dot{\alpha}}, \lambda^{\beta})$ scale as $r^0$. Note that the overall scale is immaterial for the ambitwistor variables on the boundary. Once we do this rescaling, we see that we recover the boundary incidence relations as $r\to 0$.

\subsection{Reality conditions}\label{sec.3.3}

Our considerations thus far apply to a complexified spacetime and the corresponding twistors. 
When we specialise to the case of a real slice of the spacetime, we need to impose a corresponding set of reality conditions on the twistors. We will take the ${\rm AdS}_5$ spacetime to be euclidean since we aim to focus on euclidean correlators in the dual CFT - there are likely to be additional subtleties in lorentzian signature. One of the simplifications afforded by euclidean signature is that, given a point in twistor space, the corresponding spacetime point is determined, as we will see explicitly below. See Figs. \ref{fig:fibration_on_boundary} and \ref{fig:fibration_in_bulk} for the fibration picture of this bijection. This is unlike lorentzian signature where a point in twistor space corresponds to a null ray in spacetime. 
For a discussion of reality conditions for twistors in different signatures, see for instance \cite{Adamo:2017qyl} or,  in the ${\rm AdS}_5$ context, \cite{Adamo:2016rtr}. 

In the Euclidean signature, the spacetime coordinates of complexified Minkowski space obey 
\be\label{eucl-real}
\widehat{x}^{\dot{\alpha}}_{\;\;\;\beta}=x^{\dot{\alpha}}_{\;\;\;\beta} \ ,
\ee
where $\widehat{x}^{\dot{\alpha}}_{\;\;\;\beta}$ is defined through Eq.~\eqref{hatxdef} (with bars referring to ordinary complex conjugation). Taking the complex conjugate of the boundary incidence relation  Eq.~\eqref{boundary_incidence1} and using Eq.~\eqref{eucl-real} gives 
\begin{equation} \label{incid1-conj}
    \mu^{\dot{\alpha}}=x^{\dot{\alpha}}_{\;\;\;\beta} \lambda^{\beta} \quad\Rightarrow \quad \hat{\mu}^{\dot{\alpha}}=x^{\dot{\alpha}}_{\;\;\;\beta} \hat{\lambda}^{\beta}
\end{equation}
Similarly, we also find the complex conjugate relation to Eq.~\eqref{boundary_incidence2}
\be\label{incid2-conj}
\hat{\mu}^{\dagger}_{\alpha}=-x_{\alpha}^{\;\;\; \dot{\beta}} \hat{\lambda}^{\dagger}_{\dot{\beta}}
\ee
Here we have defined the hatted twistor variables
\begin{equation}\label{hat-twstr}
\begin{split}
    &\widehat{Z}^{I}\equiv (\hat{\lambda}^{\alpha},\hat{\mu}^{\dot{\alpha}})\equiv (-\bar{\lambda}^2,\bar{\lambda}^1,-\bar{\mu}^{2},\bar{\mu}^1)\\
    & \widehat{Y}_{J}\equiv (\hat{\mu}^{\dagger}_{\beta},\hat{\lambda}^{\dagger}_{\dot{\beta}})\equiv (-\bar{\mu}_2^{\dagger},\bar{\mu}_1^{\dagger},-\bar{\lambda}_2^{\dagger},\bar{\lambda}_1^{\dagger})
    \end{split}
\end{equation}
In other words, the hatted twistors are essentially complex conjugates\footnote{Note, however, that the conjugation operation defined here is not the identity when applied twice. See below Eq.~(1.17) in \cite{Adamo:2017qyl}.} and also obey the same incidence relations as the original twistors.  

It is easy to verify that if we take the radial coordinate $r$ to be real, then we are extending the above reality conditions on the boundary into the bulk (euclidean ${\rm AdS}_5$). The reality conditions on the twistors given above also ensure that the bulk incidence relations Eq.~\eqref{bulk_incidence2} are satisfied in terms of the hatted twistors. More succinctly,
\be\label{hat-incid}
\widehat{Z}^{I} = X^{IJ} \widehat{Y}_{J} \ . 
\ee
 
The reality constraints can be used to obtain the spacetime coordinate corresponding to a point in twistor space. First we consider a point on the boundary. Since both $Z_b^I,  \widehat{Z}_b^I$ obey the incidence relation with the same $x^{\dot{\alpha}}_{\;\; \beta}$, we can use the general relation  
Eq.~\eqref{bitwstr} to write 
\begin{equation}\label{Xsol}
\begin{split}
 (X_b)^{IJ}=-\frac{Z_b^{[I}\widehat{Z}_b^{J]}}{\langle \lambda \hat{\lambda} \rangle}\quad \text{and}\quad
 (X_b)_{IJ}=-\frac{Y^b_{[I}\widehat{Y}^b_{J]}}{[\lambda^{\dagger}\hat{\lambda}^{\dagger}]} 
\end{split}
\end{equation}
Using these, the condition $(X_b)^2=0$ imposes additional constraints on the boundary ambitwistors:
\be\label{bdy-constr}
(X_b)_{IJ} (X_b)^{IJ}=0 \Rightarrow Z_b\cdot \widehat{Y}^b=\widehat{Z}_b\cdot Y^b=0
\ee
where we have used the ambitwistor condition $Z_b\cdot Y^b=0$.
From Eq.~\eqref{Xsol}, we can read off the components
\begin{align}\label{boundary_x}
    x^{\alpha\dot{\beta}}&=\frac{\hat{\mu}^{\dagger\,\alpha}\lambda^{\dagger\,\dot{\beta}}-\mu^{\dagger\,\alpha}\hat{\lambda}^{\dagger\,\dot{\beta}}}{[\hat{\lambda}^{\dagger}\lambda^{\dagger}]} \\
    &= \frac{\lambda^{\alpha}\hat{\mu}^{\dot{\beta}}-\hat{\lambda}^{\alpha}\mu^{\dot{\beta}}}{\langle\hat{\lambda}\lambda\rangle}
\end{align}
In the second relation above we have used $x^{\alpha\dot{\beta}}=x^{\dot{\beta}\alpha}$ 
We also find 
\be
x^2 = 2\frac{[\hat{\mu} \mu]}{\langle\hat{\lambda}\lambda\rangle} = 2\frac{\langle\hat{\mu}^{\dagger}\mu^{\dagger}\rangle}{[\hat{\lambda}^{\dagger}\lambda^{\dagger}]} \ .
\ee

We use a similar logic  to determine the coordinates in ${\rm AdS}_5$ given the twistors $(Z^I, Y_J)$ and their complex conjugates.  
One can write a general ansatz for $X^{IJ}$:
\begin{equation}
    X^{IJ}=\alpha \frac{Z^{[I}\widehat{Z}^{J]}}{\langle \lambda \hat{\lambda}\rangle}+\beta \epsilon^{IJKL} \frac{Y_{[K}\widehat{Y}_{L]}}{[\lambda^{\dagger}\hat{\lambda}^{\dagger}]}
\end{equation}
Note that $X^{IJ}$ is real under the conjugation operation. 
The ansatz implies
\begin{equation}
    X_{IJ}=\frac{1}{2}\alpha\epsilon_{IJKL}\frac{Z^{[K}\widehat{Z}^{L]}}{\langle \lambda \hat{\lambda}\rangle}+2\beta \frac{Y_{[I}\widehat{Y}_{J]}}{[\lambda^{\dagger}\hat{\lambda}^{\dagger}]}
\end{equation}
where we have used the identity $\epsilon_{IJKL}\epsilon^{KLMN}=2\delta^{[M}_I\delta^{N]}_J$. The bulk incidence relations contain enough information to determine $\alpha$ and $\beta$:
\begin{equation}
    \begin{split}
        Z^{I}=X^{IJ}Y_J=\alpha Z^{I}\frac{\widehat{Z}\cdot Y}{\langle \lambda\hat{\lambda}\rangle} \Rightarrow \alpha=\frac{\langle \lambda\hat{\lambda}\rangle}{\widehat{Z}\cdot Y}
    \end{split}
\end{equation}
\begin{equation}
    \begin{split}
        Y_I=-\frac{2}{r^2}X_{IJ}Z^{J}=-\frac{4}{r^2}\beta\, Y_I \frac{Z\cdot \widehat{W}}{[\lambda^{\dagger}\hat{\lambda}^{\dagger}]} \Rightarrow \beta=-\frac{r^2}{4} \frac{[\lambda^{\dagger}\hat{\lambda}^{\dagger}]}{Z\cdot\widehat{Y}}
    \end{split}
\end{equation}
where we have used $X_{IJ}X^{IJ}=2r^2$ Eq.~\eqref{rad}. 

From the conjugate incidence relation, we also find the reality constraint on the quadric 
 \begin{equation}
     \widehat{Z}^I=X^{IJ}\widehat{Y}_J=-\alpha \widehat{Z}^{I}\frac{Z\cdot \widehat{Y}}{\langle \lambda\hat{\lambda}\rangle}\Rightarrow \widehat{Z}\cdot Y=-Z\cdot \widehat{Y}.
 \end{equation}
We will denote this constraint as $\mathcal{D}$,
\be\label{bulk-constr}
\mathcal{D} \equiv \; \widehat{Z}\cdot Y+ Z\cdot \widehat{Y} =0\ .
\ee
Note that this is a weaker constraint than the one that obeyed by the boundary twistors Eq.~\eqref{bdy-constr}. 

Using these results, we get a (many-to-one - see Fig. \ref{fig:fibration_in_bulk}) correspondence between the twistors and a point in the bulk ${\rm AdS}_5$ spacetime
\begin{equation}\label{solution}
    X^{IJ} =\frac{1}{\widehat{Z}\cdot Y}\left[ Z^{[I}\widehat{Z}^{J]}+\frac{r^2}{4}\epsilon^{IJKL} Y_{[K}\widehat{Y}_{L]} \right] \ . 
\end{equation}

To find $r^2$ and $x^{\alpha\dot{\beta}}$ we look at the upper diagonal block of $X^{IJ}$ in Eq.~\eqref{bulk-pt} and equate that with r.h.s of Eq.~\eqref{solution}, which immediately gives
\begin{equation}\label{rad-sol}
    r^2=-2\frac{\widehat{Z}\cdot Y +\langle \lambda \hat{\lambda}\rangle}{[\lambda^{\dagger} \hat{\lambda}^{\dagger}]} \ .
\end{equation}
Similarly equating with any of the non-diagonal blocks in Eq.~\eqref{bulk-pt} gives
\begin{equation}\label{x-sol}
    x^{\alpha\dot{\beta}}=\frac{1}{\widehat{Z}\cdot Y} \left[ (\lambda^{\alpha}\hat{\mu}^{\dot{\beta}}-\hat{\lambda}^{\alpha}\mu^{\dot{\beta}})+\frac{r^2}{2}(\hat{\mu}^{\dagger\,\alpha}\lambda^{\dagger \, \dot{\beta}}-\mu^{\dagger\,\alpha}\hat{\lambda}^{\dagger\,\dot{\beta}})\right] \ ,
\end{equation}
which can be rewritten, using the expression of $r^2$, as
\begin{equation}
\begin{split}
    x^{\alpha\dot{\beta}}=\frac{\langle \hat{\lambda}\lambda \rangle}{\widehat{Z}\cdot Y} \left[ \frac{\lambda^{\alpha}\hat{\mu}^{\dot{\beta}}-\hat{\lambda}^{\alpha}\mu^{\dot{\beta}}}{\langle\hat{\lambda}\lambda\rangle}- \frac{\hat{\mu}^{\dagger\,\alpha}\lambda^{\dagger\,\dot{\beta}}-\mu^{\dagger\,\alpha}\hat{\lambda}^{\dagger\,\dot{\beta}}}{[\hat{\lambda}^{\dagger}\lambda^{\dagger}]}\right]+\frac{\hat{\mu}^{\dagger\,\alpha}\lambda^{\dagger\,\dot{\beta}}-\mu^{\dagger\,\alpha}\hat{\lambda}^{\dagger\,\dot{\beta}}}{[\hat{\lambda}^{\dagger}\lambda^{\dagger}]} \ .
    \end{split}
\end{equation}
On the boundary, this is consistent with the expression for $x^{\alpha\dot{\beta}}$ which we obtained in Eq.~\eqref{boundary_x} since the first term in square brackets above vanishes. As another consistency  check, we can reproduce (\ref{bulk_incidence2}) with these expressions of $r^2$ and $x^{\alpha \dot{\beta}}$. For example 
\begin{equation}\label{consistency1}
    x^{\alpha\dot{\beta}}\lambda^{\dagger}_{\dot{\beta}}=(\lambda^{\alpha}-\mu^{\dagger\,\alpha})-\frac{1}{\widehat{Z}\cdot Y}(\hat{\lambda}^{\alpha}\mu^{\dot{\beta}}\lambda^{\dagger}_{\dot{\beta}}+\lambda^{\alpha}\hat{\lambda}^{\beta}\mu^{\dagger}_{\beta}+\mu^{\dagger\,\alpha}\langle \lambda \hat{\lambda}\rangle)
\end{equation}
where we have used the expression of $r^2$. Now using the ambi-twistor condition, third and fourth terms of the above expression become
\be 
\hat{\lambda}^{\alpha}\mu^{\dot{\beta}}\lambda^{\dagger}_{\dot{\beta}}+\lambda^{\alpha}\hat{\lambda}^{\beta}\mu^{\dagger}_{\beta}=(-\hat{\lambda}^{\alpha}\lambda^{\beta}+\lambda^{\alpha}\hat{\lambda}^{\beta}) \mu^{\dagger}_{\beta}=-\epsilon^{\alpha\beta}\langle \lambda \hat{\lambda} \rangle \mu^{\dagger}_ {\beta} 
\ee
This cancels the last term in (\ref{consistency1}) and we get back the first equation in (\ref{bulk_incidence2}).

\section{Twistor space covering maps in ${\rm AdS}_5$}\label{sec.4}

We have described the twistor space geometry of ${\rm AdS}_5$ and the reality conditions which fix a point in the bulk in terms of the twistor variables. We want to use this geometrical picture to describe string configurations which capture the dual ${\cal N}=4$ super Yang-Mills theory as recently proposed in \cite{Gaberdiel:2021qbb, Gaberdiel:2021jrv}. We will not give a complete picture here -- this will be described elsewhere -- but instead focus on a special configuration which is a direct generalisation of the ${\rm AdS}_3$ case as described in Sec. \ref{sec.3}.  

The basic idea is that we will consider holomorphic worldsheet twistor fields $Y^I(z), Z_I(z)$, together with the corresponding right moving antiholomorphic twistor fields $\widehat{Z}^{I}(\bar{z}), \widehat{Y}_{J}(\bar{z})$, which are related to the holomorphic fields by the conjugation operation defined in Eq.~\eqref{hat-twstr}. Here the conjugation is assumed to act on the worldsheet variables on which the fields now depend, in the standard way $z \to \bar{z}$. Thus the right movers are not independent degrees of freedom and determined in terms of the left movers by Eq.~\eqref{hat-twstr}. This is consistent with the observation in \cite{Gaberdiel:2021qbb, Gaberdiel:2021jrv} that we only need one copy of the wedge modes (i.e. not both left and right movers) of the twistor fields to account for the correct spectrum of the free super Yang-Mills theory\footnote{It may be that we only need to impose this reality condtion on the finite number of wedge modes of the worldsheet twistor fields. Since these are the only ones excited in the classical solutions, we will not be able to distinguish the two possibilities in the present discussion. The right prescription will presumably arise from a complete worldsheet analysis. We thank Matthias Gaberdiel for discussions on this point and ongoing collaborations \cite{GGKM}.}. 

To connect with the Feynman diagrams for the dual gauge theory operators, we will need to translate twistor string configurations into ${\rm AdS}$ spacetime configurations. We will determine the latter by requiring that the twistors obey the ${\rm AdS}_5$ twistor incidence relations, together with the reality conditions. In other words, we now require the stringy versions of Eq.~\eqref{bulk_incidence2}. 
\be\label{stringy-bulk}
\begin{split}
        \mu^{\dagger}_{\alpha}(z)+X_{\alpha}^{\;\;\;\dot{\beta}}(z, \bar{z})\lambda^{\dagger}_{\dot{\beta}}(z) &= \lambda^{\beta}(z) \epsilon_{\beta \alpha}\\
        \mu^{\dot{\alpha}}(z)-X^{\dot{\alpha}}_{\;\;\; \beta}(z, \bar{z})  \lambda^{\beta}(z) &=\frac{1}{2}R^2(z, \bar{z}) \epsilon^{\dot{\alpha}\dot{\beta}} \lambda^{\dagger}_{\dot{\beta}}(z) \ .
    \end{split}
\ee
Here we have denoted the string configurations in ${\rm AdS}_5$ by $X^{\dot{\alpha}}_{\;\;\; \beta} \lambda^{\beta}(z, \bar{z})$ and the radial $R(z, \bar{z})$. As we will see shortly, unlike in 
${\rm AdS}_3$, the longitudinal coordinates $X^{\dot{\alpha}}_{\;\;\; \beta}$ are now no longer purely holomorphic in the worldsheet coordinates. Ultimately, these relations will not need to be postulated; rather, we expect they will arise from an analysis of the ward identities of the worldsheet CFT as in \cite{Dei:2020zui}. This is currently under investigation \cite{GGKM}. 

In fact, since the bulk incidence relations Eq.~\eqref{stringy-bulk} hold pointwise on the worldsheet, we will have the same expressions for the bulk coordinates $R(z, \bar{z}),  X^{\dot{\alpha}}_{\;\;\; \beta}(z, \bar{z})$, in terms of the twistor fields, as given in Eqs.~\eqref{rad-sol}, \eqref{x-sol}. Since these depend on both the twistor fields and their conjugates, we clearly see that they cannot be holomorphic (or anti-holomorphic). Note that since the twistor fields obey the bulk incidence relation it also immediately follows that they obey, pointwise, the ambitwistor constraint 
\be\label{ambconstr}
\lambda^\dag_{\dot{\alpha}}(z)\mu^{\dot{\alpha}}(z)+\mu^\dag_{\alpha}(z)\lambda^{\alpha}(z) =0 \ .
\ee
One may alternately view the ambitwistor constraint as primary and then the requirement of the bulk incidence relations in Eq.~\eqref{stringy-bulk} follows. The latter are a general way of satisfying the ambitwistor constraints -- see the discussion below Eq.~\eqref{bulk_incidence2}. The ambitwistor constraint arose as a fundamental gauge constraint requirement in the worldsheet proposal of 
\cite{Gaberdiel:2021qbb, Gaberdiel:2021jrv}.

We will further assume that as in ${\rm AdS}_3$, the worldsheet is localised near the boundary i.e. that $R(z, \bar{z}) \propto \epsilon \to 0$. This is also physically motivated by the fact that the dual free theory is at the UV fixed point and hence essentially glued to the boundary of ${\rm AdS}_5$, using the relation between scale and the radial direction. We expect, however, there to be a nontrivial radial (Liouville) profile $\partial{\ln{R}}$ which we will not determine here.    

As discussed in Sec. \ref{sec.3.2}, when the radial coordinate goes to zero, the bulk incidence relations reduce to the boundary incidence relations. Thus for string configurations near the boundary we must have the analogues of Eqs~\eqref{boundary_incidence1},\eqref{boundary_incidence2}, namely, 
\be\label{stringy-bdy}
\begin{split}
\mu^{\dot{\alpha}}(z)&=X^{\dot{\alpha}}_{\;\;\;\beta}(z, \bar{z}) \lambda^{\beta}(z)  \\
\mu^{\dagger}_{\alpha}(z) &=-X_{\alpha}^{\;\;\; \dot{\beta}}(z, \bar{z}) \lambda^{\dagger}_{\dot{\beta}}(z) \ .
\end{split}
\ee
We will mostly work with these (stringy) incidence relations on the boundary given our physical assumption of the worldsheet being localised there. 

 As mentioned at the beginning of the section, we are also imposing the reality conditions pointwise in $(z, \bar{z})$. Therefore we can use the above incidence relations and the corresponding conjugate relations to obtain the stringy generalisation of Eq.~\eqref{boundary_x}
 \begin{align}\label{boundary_string}
  X^{\alpha\dot{\beta}}(z, \bar{z})&= \frac{\lambda^{\alpha}(z)\hat{\mu}^{\dot{\beta}}(\bar{z})-\hat{\lambda}^{\alpha}(\bar{z})\mu^{\dot{\beta}}(z)}{\langle\hat{\lambda}(\bar{z})\lambda(z)\rangle} \\
    &= \frac{\hat{\mu}^{\dagger\,\alpha}(\bar{z})\lambda^{\dagger\,\dot{\beta}}(z)-\mu^{\dagger\,\alpha}(z)\hat{\lambda}^{\dagger\,\dot{\beta}}(\bar{z})}{[\hat{\lambda}^{\dagger}(\bar{z})\lambda^{\dagger}(z)]}
\end{align}
 One sees, somewhat reassuringly, that the spacetime configurations $X^{\alpha\dot{\beta}}(z, \bar{z})$ are not purely holomorphic or anti-holomorphic even though the twistor fields are. Nevertheless, the holomorphic twistor fields obey the incidence relations with $X^{\alpha\dot{\beta}}(z, \bar{z})$ in Eq.~\eqref{stringy-bdy}, as can be seen from the explicit form in Eq.~\eqref{boundary_string}.  
 
The mixed $z$ dependence here is unlike the ${\rm AdS}_3$ case where there was a split of the boundary spacetime coordinates into what we had denoted as $\Gamma(z)$ and its conjugate. 
Despite that, there is a sense in which there is a kind of local holomorphy in the boundary spacetime. 
This is reflected in the fact that even though $X^{\alpha\dot{\beta}}(z, \bar{z})$ is not holomorphic, the matrix $\bar{\partial}X^{\alpha\dot{\beta}}(z, \bar{z})$ (as also $\partial X^{\alpha\dot{\beta}}(z, \bar{z})$) has a zero eigenvector. 

This follows directly by applying $\bar{\partial}$ on both sides of the two equations in Eq.~\eqref{stringy-bdy}. We obtain
\be 
 \begin{split}
\bar{\partial}X^{\dot{\alpha}}_{\;\;\;\beta}(z, \bar{z}) \lambda^{\beta}(z) &= 0 \ , \\
\bar{\partial}X_{\alpha}^{\;\;\; \dot{\beta}}(z, \bar{z}) \lambda^{\dagger}_{\dot{\beta}}(z) &= 0\ .
\end{split}
\ee
In other words, we see that $\lambda^{\beta}(z)$ is the zero eigenmode of $\bar{\partial}X^{\dot{\alpha}}_{\;\;\;\beta}(z, \bar{z})$ while $\lambda^{\dagger}_{\dot{\beta}}(z)$ is the zero eigenvector of the transposed matrix  $\bar{\partial}X_{\alpha}^{\;\;\; \dot{\beta}}$. 
One immediate consequence of this, as can also be easily verified directly using the expressions in Eq.~\eqref{boundary_string}, is that 
\be\label{nilp}
\bar{\partial}X_{\alpha\dot{\beta}}\bar{\partial}X^{\alpha\dot{\beta}} =0 \ .
\ee
For $\partial X^{\alpha\dot{\beta}}$, we can similarly write down the zero eigenvectors in terms of the conjugate twistor fields and the analogue of Eq.~\eqref{nilp}. We interpret the presence of this zero eigenvector of $\bar{\partial}X^{\dot{\alpha}}_{\;\;\;\beta}(z, \bar{z})$ as indicating that, locally, on the worldsheet and hence in space time, we can always view the string configuration $X^{\dot{\alpha}}_{\;\;\;\beta}(z, \bar{z})$ as a holomorphic embedding into the boundary of an ${\rm AdS}_3$ subspace of the bulk spacetime. We will exploit this interpretation in what follows. 

We also note that if we consider the more general stringy incidence relations, we can arrive at a similar conclusion. Namely, applying $\bar{\partial}$ on the first equation of Eq.~\eqref{stringy-bulk} we find that $\lambda^{\dagger}_{\dot{\beta}}(z)$ continues to be the zero eigenvector of $\bar{\partial}X_{\alpha}^{\;\;\; \dot{\beta}}(z, \bar{z})$. However, we see from the non-holomorphic dependence of $R^2(z, \bar{z})$ in the RHS of the second equation of  Eq.~\eqref{stringy-bulk} that $\lambda^{\beta}(z)$ is no longer the zero eigenmode of $\bar{\partial}X^{\dot{\alpha}}_{\;\;\;\beta}(z, \bar{z})$. 

\subsection{Maps to an ${\rm AdS}_3$ subspace}

In this subsection, we will consider the subset of solutions where the string lies entirely within an ${\rm AdS}_3$ subspace, or equivalently the boundary insertion points $x_i$ lie in a two dimensional plane. Note that this can always be done for the case of two and three point functions. For four point functions and higher this is a special configuration. 
We will work in this special kinematic setup as it will enable us to be very explicit. We will also be able to illustrate, in the next section, how the corresponding covering maps reproduce the boundary free field Feynman diagram answers. 

Since $\bar{\partial}X^{\dot{\alpha}}_{\;\;\;\beta}(z, \bar{z})$ has one zero eigenvalue, we will choose it to be in diagonal form (which can always be done locally, but here we are doing so globally)
\be\label{diagdelx}
\bar{\partial}X^{\dot{\alpha}}_{\;\;\;\beta}(z, \bar{z})=
\begin{bmatrix}
        0&& 0 \\
        0 && -\bar{\partial}\bar{V}(\bar{z}) 
\end{bmatrix} \ .
\ee
Recall from the general form of $X^{\dot{\alpha}}_{\;\;\;\beta}$ (see Eq. (\ref{Z-Y})) that this implies we have taken $\bar{\partial}U = \bar{\partial}\bar{U}=0$ i.e. $U=0$ without loss of generality. We also have $\bar{\partial}V=0$ i.e. $V=V(z)$ and thus the purely antiholomorphic dependence in $\bar{V}(\bar{z})$. In other words, we have the string configuration
\be\label{diagx}
X^{\dot{\alpha}}_{\;\;\;\beta}(z, \bar{z})=
\begin{bmatrix}
        -V(z) && 0 \\
        0 && -\bar{V}(\bar{z}) 
\end{bmatrix} \ .
\ee
As mentioned, the string configuration is entirely in the $(x_1,x_2)$ plane.  

The zero eigenvector $\lambda^{\beta}(z)$ of Eq.~\eqref{diagdelx} then has the form
\be\label{diagev}
\lambda^{\beta}(z) =
 \begin{bmatrix}
        \lambda^{1}(z)\\ 0
 \end{bmatrix} \ .
\ee
Together with Eq.~\eqref{diagx} and Eq.~\eqref{stringy-bdy} we have 
 \be\label{diagmu}
 \mu^{\dot{\alpha}}(z) = -
 \begin{bmatrix}
       V(z) \lambda^{1}(z)\\ 0
 \end{bmatrix} \ .
\ee

Thus $\mu^1(z)=-V(z)\lambda^1(z)$. Since $V(z)$ is a (finite degree) covering map from the genus zero worldsheet to the $S^2$ boundary of the ${\rm AdS}_3$, it will be a ratio of polynomials. As in the case of ${\rm AdS}_3$, we can achieve this if the twistor fields are given by rational functions. In other words, we are only exciting finitely many modes around each worldsheet insertion of the vertex operators. 
We will therefore consider fields of the form
\be\label{4dtwstrs} 
\lambda^1(z) = \frac{R_{n-1}(z)Q^1_{N}(z)}{\prod_{i =1}^n (z- z_i)^{\frac{w_i}{2}}} \ , \, \, \,  
\mu^1(z) = \frac{R_{n-1}(z)P^1_{N}(z)}{\prod_{i =1}^n (z- z_i)^{\frac{w_i}{2}}} \ .
\ee 
This is a generalisation of the expressions Eqs.~\eqref{twstrQ},\eqref{twstrP}. We note that the order of the poles in the denominator in the above is shifted. If $w^{2d}_i, w^{4d}_i$ denote the spectral flow parameters of the ${\rm AdS}_3$ and ${\rm AdS}_5$ theories, (or equivalently, the twisted sector and the number of Yang-Mills bits in the dual CFT) respectively, then we have a shift (see Eq.~(5.11) of \cite{Gaberdiel:2021jrv}) 
\be\label{2d4d}
w^{4d}_i=  w^{2d}_i+1 \ .
\ee
At the risk of creating confusion we will drop the superscripts, with the context hopefully making clear which label is being considered and Eq.~\eqref{2d4d} being the dictionary between the two.  

We see from Eq.\eqref{4dtwstrs} that the twistor fields $\lambda^1(z), \mu^1(z)$ near $z= z_i$ have a singularity of order $\frac{1}{(z-z_i)^{\frac{w}{2}}}$. This is consistent with the fact that the twistors have spin half and only have wedge modes with mode number $r\leq \frac{w_i-1}{2}$ excited. At the same time we expect a branching behaviour of order $(w_i-1)$ near each such vertex operator insertion since $r\geq - \frac{w_i-1}{2}$ (cf. the shift of Eq~\eqref{2d4d})\footnote{This is true for a generic vertex operator. However, for the BPS highest weight state the branching is of order $w_i$ due to additional mode annihilation. We thank M. Gaberdiel for bringing this to our attention. \label{fn9}}. 
In other words, from Eq.~\eqref{4dtwstrs}, we can define the covering map in position space
\be\label{4dcover}
-V(z) =\frac{\mu^1(z)}{\lambda^1(z)} = \frac{P_N(z)}{Q_N(z)} \ .
\ee
We expect that $V(z) \sim V_i +a_i(z-z_i)^{w_i-1}$, where $V_i= (x_{(i)1}+ix_{(i)2})$. 

Note also that we have the condition $N+n-1=\frac{1}{2}\sum_i w_i$ for the fields in Eq.~\eqref{4dtwstrs}  to be rational functions on the worldsheet with the prescribed poles. Therefore, $\frac{1}{2}\sum_i w_i$ must be an integer. Indeed this is consistent with the fact that the number of free field wick contractions in a Feynman diagram for a correlator built of operators with $\{w_i\}$ fields, $\langle \prod_{i=1}^n{\cal O}^{(w_i )}(x_i)\rangle$, is given precisely by $\frac{1}{2}\sum_i w_i$. Finally, since the branching behaviour for a $w_i$ spectrally flowed operator is $(w_i-1)$, the degree $N$ of the polynomials, $Q$ and $P$ are given by the Riemann-Hurwitz formula as in ${\rm AdS}_3$ with $N$ being the nett degree of the branched cover. The polynomial $R_{n-1}(z)$ is a common overall factor in Eq.~\eqref{4dtwstrs} and does not affect the branching behaviour. 

Thus when the points $x_i$ are all in a plane, the twistor solutions are a generalisation of the ones we saw in $AdS_3$. The main difference is that for a string at the boundary, the incidence relations for $AdS_5$ Eq.~\eqref{stringy-bulk} reduce to Eq.~\eqref{stringy-bdy}. Hence the analogue of the terms on the RHS of Eq.~\eqref{incid1} are absent here. As a result, in Eq.~\eqref{4dtwstrs} we do not have the analogue of the second term on the RHS of Eq.~\eqref{Pdef}. 

We have not discussed the twistor fields $\lambda^{\dagger}_{\dot{\beta}}(z)$ thus far. For the choice of $\bar{\partial}X^{\dot{\alpha}}_{\;\;\;\beta}(z, \bar{z})$ in Eq.~\eqref{diagdelx}, $\lambda^{\dagger}_{\dot{\beta}}(z)$ is also proportional to the vector in 
Eq~\eqref{diagev}, with a non-zero component $\lambda^{\dagger}_{1}(z)$. This is because it is a zero eigenvector of the transpose of the matrix, $\bar{\partial}X^{\dot{\alpha}}_{\;\;\;\beta}(z, \bar{z})$. However, given the second incidence relation in Eq.~\eqref{stringy-bdy}, we see that $\mu^{\dagger}_{1}(z)= V(z)\lambda^{\dagger}_{1}(z)$. Since the branching behaviour at the insertions, which is determined by the spectral flow, is the same for both sets of twistor fields $(\lambda^{\dagger}_{\dot{\beta}}(z), \lambda^{\beta}(z))$, this implies that they are determined by the same covering map. Thus the same polynomials $P_N(z)$ and $Q_N(z)$ determine $\mu^{\dagger}_{1}(z)$ and $\lambda^{\dagger}_{1}(z)$ respectively. Therefore these are not independent degrees of freedom. 

We expect the general solution to be determined by double the number of degrees of freedom of the special kinematic setup. In other words, there will now be two $\lambda^{\beta}(z)$ and the corresponding two $\mu^{\dot{\alpha}}(z)$. And we expect the $\lambda^{\dagger}_{\dot{\beta}}(z)$ and $\mu^{\dagger}_{\alpha}(z)$ not to be independent of the former. This will be studied elsewhere \cite{GGKM}. 
In the next section, we will use the restricted covering maps of this section to show how it reproduces the Yang-Mills propagators in a very natural way which parallels that in the $AdS_3/CFT_2$ case \cite{Gaberdiel:2020ycd}.

\section{Relation to Feynman diagrams in gauge theories}\label{sec.5}

Consider the correlation function of $n$ gauge invariant operators $\mathcal{O}^{(w_i)}(x_i)$ (made out of $w_i$ `letters' or singletons) in  free $SU(N)$ Yang-Mills theory, in the large $N$ limit. For concreteness, we will assume we are working with ${\cal N}=4$ SYM and the $\mathcal{O}^{(w)}$ are built from $w$ scalar fields. Then   
\begin{equation}\label{feyncorr}
    G_n(\{x_i\})=\langle \mathcal{O}^{(w_1)}(x_1)\cdots \mathcal{O}^{(w_n)}(x_n) \rangle_{planar} = \sum_{\{n_{ij}\}}C_{\{n_{ij}\}} \prod_{(i,j)}\left(\frac{1}{x_{ij}^2}\right)^{n_{ij}}
\end{equation}
is simply given in terms of sums of products of the individual propagators $\left(\frac{1}{x_{ij}^2}\right)^{n_{ij}}$, where $n_{ij}=n_{ji}$ denote the number of (homotopic) Wick contractions between the pair of vertices at positions $(x_i, x_j)$. Here $C_{\{n_{ij}\}}$ are combinatorial factors counting various planar diagrams, whose precise form is inessential for what we are about to describe. 
We also have the constraints on the $\{n_{ij}\}$ 
\be\label{vertx-constr}
\sum_{j}n_{ij} =w_i \ ,
\ee
holding for each vertex $(i)$. Each of the contributions in Eq.~\eqref{feyncorr} corresponds to a planar Feynman graph with $n$ vertices and $n_{ij}$ homotopic edges between the pair $(ij)$ of vertices. Thus there are a total of $\frac{1}{2}\sum_{i,j}n_{ij}=\frac{1}{2}\sum_iw_i$ edges in the graph. 

In ${\rm AdS}_3$, the correlator of twisted sector fields corresponding to Eq.~\eqref{feyncorr} was  given by a sum over contributions associated to different covering maps in the Lunin-Mathur description \cite{Lunin:2001}. These, in turn were associated \cite{Pakman:2009} in a one to one way with a set of Feynman diagrams which captured the covering map data, as briefly recalled in Appendix \ref{appB}. Here we want to do something parallel, but in the opposite direction, in a sense. Namely, we would like to associate a covering map from the worldsheet for each familiar Feynman diagram contribution. We will see that, at least for the restricted kinematic configurations of the previous subsection, we will have an analogous picture\footnote{As mentioned, placing all the $n$-points $(x_i)$ in a 2d plane is always possible for $n=2,3$, so this is really a special choice only for four and higher point functions.}. Each graph will therefore correspond to a certain point in the moduli space of the closed string theory (which admits this covering map with specified branching data). We will see, via a connection of the Schwarzian of this covering map to the unique Strebel differential at that point in the moduli space, that this association of Feynman graphs to closed string worldsheets is as per the Strebel prescription of \cite{Gopakumar:2005fx, Razamat:2008zr}. This is also in parallel to the $AdS_3/CFT_2$ case as in \cite{Gaberdiel:2020ycd} -- see Appendix \ref{appB} and specifically Fig. \ref{fig:flowchart}. And very strikingly, the weight, associated to each Feynman diagram, coming from the propagators in Eq.~\eqref{feyncorr} can be reproduced in terms of the area of the worldsheet in the so-called Strebel gauge \cite{Gopakumar:2011ev}, again in parallel to the ${\rm AdS}_3$ case \cite{Gaberdiel:2020ycd}. Let us now describe how this comes about. 

\subsection{Two point function}

We start with the simplest correlator, a two point function. Then we can always choose the two points $(x_i, x_j)$ to lie in a specified plane on the boundary of the ${\rm AdS}_5$. Since we are considering scalar fields, we can take these to be the highest weight BPS state in the SYM theory corresponding to the $w$-spectrally flowed vacuum sector $|0\rangle_w$ \cite{Gaberdiel:2021qbb, Gaberdiel:2021jrv} (together with the conjugate field). We will view the corresponding states on the worldsheet to be inserted at $z=(0, \infty)$, respectively, without loss of generality. For such a correlator, as discussed in the previous section, the classical twistor field will have branching of order $w$ at $z=(0, \infty)$ (see footnote \ref{fn9}). The corresponding covering map in Eq.~\eqref{4dcover} is then fixed (after absorbing a constant into a rescaling of $z$) to take the form 
\be\label{2pt-cov}
V(z) = \frac{V_j \, z^{w}+V_i}{z^{w}+1} \ ,
\ee
where $Z_i$ is the complex coordinate in the $(1,2)$ plane corresponding to $x_i$ (see below Eq.~\eqref{4dcover}). This corresponds to the polynomials in Eq.~\eqref{4dtwstrs} and Eq.~\eqref{4dcover} taking the values
\be
P_{w}(z)= V_j \, z^{w}+V_i  \, \,; \,\,\,\,\,\, Q_{w}(z) = z^{w}+1 \ .
\ee
From the point of the view of the twistor fields Eq. \eqref{4dtwstrs}, taking into account the weight half of the spinor fields, we see that we are exciting only the wedge modes with $r= \pm \frac{(w-1)}{2}$ as appropriate for the highest weight BPS state (and its conjugate). 

As we will shortly see, it will be convenient to view the covering map in coordinates $z=e^{2\pi i \tfrac{u}{w}}$ which maps the vertical strip $(0< {\rm Re} \,u \leq w)$ onto the sphere such that $z=(0, \infty)$ are images of  $u=\pm i\infty$, respectively, on the strip. In terms of these coordinates the covering map takes the form 
\be\label{tanmap}
\Gamma(u) \equiv V(z(u)) = \frac{V_i+V_j}{2}+\frac{V_i-V_j}{2i}\,\tan(\pi u) \ .
\ee
In this form, we now observe that this is essentially the unique map for which the Schwarzian is a constant:
\be
S[\Gamma(u)]= \frac{\Gamma'''}{\Gamma'} - \frac{3}{2} \Bigl( \frac{\Gamma''}{\Gamma'} \Bigr)^2 =2\pi^2 \ .
\ee 
At the same time, we know that the unique Strebel quadratic differential on the strip with poles only at $u=\pm i\infty$ is also just $du^2$.  So we see that, as in \cite{Gaberdiel:2020ycd} (cf. Eq.~(6.5) there), the Strebel differential on the worldsheet is identified with the Schwarzian of the covering map. 
\be
\phi_S(u) du^2 = \frac{1}{2\pi^2}S[\Gamma(u)]du^2 \ .
\ee
Note that the Schwarzian also transforms as a quadratic differential and hence this is a coordinate independent statement. Indeed, in terms of the coordinate $z$, we have the Strebel differential 
\be\label{schwz-strb}
du^2= -\frac{w^2}{4\pi^2} \frac{dz^2}{z^2} \ ,
\ee
which has double poles, as expected at $z=0, \infty$ with ``residues" $\propto w$. 

This identification enables us to interpret this worldsheet as that corresponding to the Feynman diagram associated to this two point function. Then the strip of width $w$ is nothing other than the $w$ double line edges glued together. This Strebel prescription of \cite{Gopakumar:2005fx, Razamat:2008zr, Gopakumar:2011ev} was seen to be also realised in the large $w_i$ limit in the case of ${\rm AdS}_3/CFT_2$ in the tensionless regime. 

\begin{figure}[!htb]
    \centering
   \begin{tikzpicture}[scale=0.85]
   \draw[color=white, fill=cyan!10] (0,0-1.5)--(1.5,0-1.5)--(1.5,5+1.5)--(0,5+1.5)--(0,0-1.5);
    \draw[color=white, fill=cyan!23] (1.5,0-1.5)--(3,0-1.5)--(3,5+1.5)--(1.5,5+1.5)--(1.5,0-1.5);
\draw[very thick, blue] (0,-1.5)--(0,5+1.5) (1.5,0-1.5)--(1.5,5+1.5) (3,0-1.5)--(3,5+1.5) (7.5,0-1.5)--(7.5,5+1.5);
\draw[dashed] (0,0.2-1.5)--(7.5,0.2-1.5) (0,4.8+1.5)--(7.5,4.8+1.5);
\fill[black] (0.75+3,2.5) circle (0.05) (0.75+3.5,2.5) circle (0.05) (0.75+4,2.5) circle (0.05) (0.75+4.5,2.5) circle (0.05) (0.75+5,2.5) circle (0.05) (0.75+5.5,2.5) circle (0.05) (0.75+6,2.5) circle (0.05) (0.75+6.5,2.5) circle (0.05);
\fill[black] (3.75,-1-1.5) circle (0.08) (3.75,6+1.5) circle (0.08);
\node at (8.1,0.2-1.5) { $\boldsymbol{-\,iL}$};
\node at (8.1,4.8+1.5) { $\boldsymbol{+\,iL}$};
\node at (4.2,-1-1.5) { $\large\boldsymbol{z_j}$};
\node at (4.2,6+1.5) { $\large \boldsymbol{z_i}$};
\draw[<-] (0,2)--(3.6,2);
\node at (4.3,2) {$w$};
\draw[->] (5,2)--(7.5,2);
\end{tikzpicture}
    \caption{A vertical strip of width $w$ in the $u$-plane. Regulators at $u=\pm iL$ are shown.}
    \label{fig:w-strips}
\end{figure}
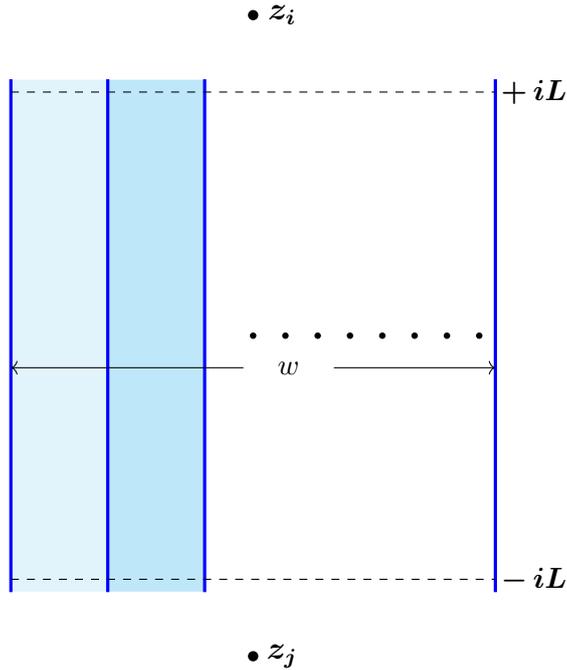

We can go further and look at the Nambu-Goto area of the worldsheet in the metric induced by the Strebel differential. This had also entered into the weight associated with the correlator in ${\rm AdS}_3$ \cite{Gaberdiel:2020ycd}. The area is simplest to compute in the $u$ coordinates. It is just the area of the strip. The width is proportional to $w$. However, the length is formally infinite. This is a reflection of the UV divergence of the field theory. To exhibit this, we put cutoffs 
$-L \leq {\rm Im} \, u \leq L$ with $L \gg 1$. We now denote   
\be\label{epsdef}
|V_i - \Gamma(u= iL)| = |V_j- \Gamma(u=-iL)| = \epsilon  
\ee
where $\epsilon$ is a short distance cutoff in spacetime. Using the behaviour 
\begin{equation}
           i\tan{u}=\mp(1-2\,e^{\pm 2iu}), \quad u\to \pm i\infty ,
\end{equation}
we find for the covering map Eq.~\eqref{tanmap} that
\begin{equation}
        \epsilon = |V_i - \Gamma(u= iL_i)|  = |V_i-V_j|\exp(-2\pi L),\quad L\rightarrow \infty \ .
\end{equation}

In other words, 
\be
L = \frac{1}{4\pi}\ln {\left(\frac{x_{ij}^2}{\epsilon^2}\right)} \ ,
\ee
where we have used $|V_i-V_j|^2= (x_{ij})^2$. 
We then see that the (regulated) area of the strip in Strebel gauge is $A_S= 2Lw$. Hence, the natural Nambu-Goto weight
\be\label{streb-prop1}
e^{-2\pi A_S} = \epsilon^{2w} \left(\frac{1}{x_{ij}^2}\right)^w \ .
\ee
We reproduce the propagator of the free gauge theory simply via the area of the string worldsheet in the special Strebel gauge we are working in. Note that the $\epsilon$ dependence in the answer is just a multiplicative renormalisation that can be absorbed as usual.

\subsection{Multi point correlator}

We have seen that for the two point function, we can reproduce the correct propagator Eq.~\eqref{streb-prop1} from the Strebel area of the Strebel diffferential, given by the Schwarzian of the covering map Eq.~\eqref{2pt-cov} as in Eq.~\eqref{schwz-strb}. 
To generalise to a multi-point correlator, we will use the results of \cite{Gaberdiel:2020ycd}, restricting to the special kinematic configuration where all the $n$-points are in the same plane. In the regime of large $w_i$, which we are considering, the covering maps with specified branching data $(x_i, w_i)$ were explicitly characterised in terms of a spectral curve for a Penner-like matrix model. The spectral curve, and thus the covering map, was determined in terms of a set of integers $n_{ij}$ obeying the constraint Eq. ~\eqref{vertx-constr}. Each such covering map was associated with a Feynman diagram 
with $n$ vertices with $w_i$ double lines emanating from each of the vertices. The $n_{ij}$ were the number of edges between the pair of vertices $(i, j)$. 

Furthermore, it was argued in \cite{Gaberdiel:2020ycd} that to leading order (i.e. for large $w_i$), this spectral curve, given by the Schwarzian of the covering map, was, rather remarkably, proportional to the unique Strebel differential on the worldsheet with specified Strebel lengths $n_{ij}$.
The Feynman-'t Hooft diagram associated to the covering map was then simply the Strebel construction of the closed string worldsheet, with $n$ punctures, formed from from gluing strips together as in the general prescription of \cite{Gopakumar:2005fx, Razamat:2008zr, Gopakumar:2011ev}. 

\begin{figure}[!htb]
    \centering
    \vspace{3mm}
    \begin{tikzpicture}[scale=0.85]
\draw[color=white, fill=cyan!10] (0, 0)--(3.92,0.694)--(7,-1)--(3.88,-1.84)--(0,0);
\draw[ultra thick, blue] (0.49, 0.087)--(3.92,0.694) (0.485,-0.23)--(3.88,-1.84) (0.25,0.43)--(2,3.46) (-0.25,0.43)--(-2,3.46) (-0.43,-0.25)--(-3.44,-2) (0.25,-0.43)--(2,-3.46);
\draw[dashed, very thick, blue] (3.92,0.694)--(7,-1) (3.88,-1.84)--(7,-1);

\draw [<->, thick] (0.965,-0.4) arc (-15:1:2cm);
\fill[black] (0,0) circle (0.1) (7,-1) circle (0.1);
\node at (1.99,-0.2) {$\boldsymbol{2\pi \frac{n_{ij}}{w_i}}$}; 
\node at (-0.5,0) {$\large \boldsymbol{z_i}$};
\node at (7.5,-1) {$\large \boldsymbol{z_j}$};
\end{tikzpicture}
    \caption{The local neighbourhood near a double pole of the Strebel differential - where strips such as in Fig. \ref{fig:w-strips} are glued together. Compare with the middle picture of Fig. \ref{fig:triptych}.}
    \label{fig:Gluing}
\end{figure}
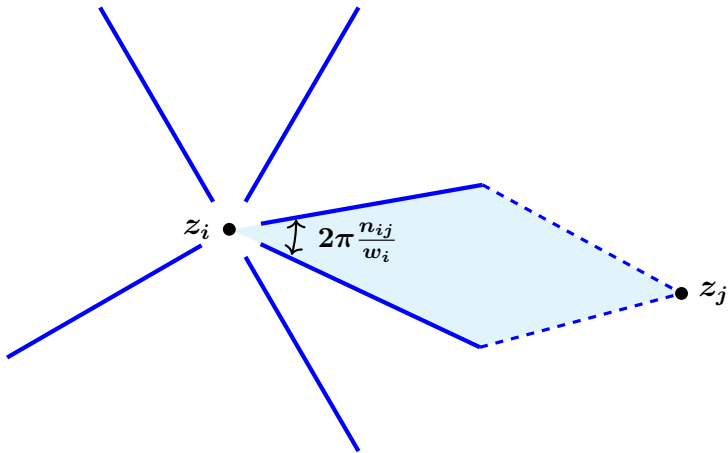

All these considerations go through in our present context of a restricted kinematic configuration. We will therefore use the connection between the Strebel differential and the Schwarzian of the covering map to locally describe the covering map. Thus, we will consider the worldsheet, which admits a covering map, to be comprised of vertical strips of width $n_{ij}$ which connect the pair of worldsheet points $(z_i, z_j)$. The Strebel differential has no poles or zeroes on this strip, except at the boundaries and hence we can find a local coordinate $u_{(ij)}$ on this strip in which it is just $du_{(ij)}^2$. The Schwarzian of the associated covering map $\Gamma(u_{(ij)})$ is a constant. Given that the covering map must interpolate (as $u_{(ij)}\to \pm i\infty$) between the space time points with planar coordinates $V_i$ and $V_j$, this fixes the covering map in this coordinate chart to be as in Eq.~\eqref{tanmap}. 
\be\label{tanmap2}
\Gamma(u_{(ij)})  = \frac{V_i+V_j}{2}+\frac{V_i-V_j}{2i}\,\tan(\pi u_{(ij)}) \ .
\ee
Note that this is the covering map on a local patch of the Riemann surface. To obtain the full covering map, one will have to glue together these maps in different patches. There is a Strebel differential $\phi_S(z)dz^2$ (with given Strebel lengths $n_{ij}$) which takes the form $(\frac{du_{(ij)}(z)}{dz})^2dz^2$ in each strip and 
\be
\phi_S(z)dz^2 \sim -\frac{w_i^2}{4\pi^2} \frac{dz^2}{(z-z_i)^2} 
\ee
in the vicinity of the insertions $z_i$. 
The covering map is then obtained by gluing together the different patches $u_{(ij)}(z)$. See fig. \ref{fig:Gluing}.

However, for the purpose of obtaining the Strebel area, we will not really need the global covering map or Strebel differential. We simply add up the areas of each of these strips, appropriately regularised. 
Given that in each strip the covering map is given by Eq.~\eqref{tanmap2}, we just repeat the considerations of the previous subsection. We thus obtain a regularised strip of width $n_{ij}$ - the Strebel length - and vertical size 
\be
2L_{ij} = \frac{1}{2\pi}\ln {\left(\frac{x_{ij}^2}{\epsilon^2}\right)} \ ,
\ee
with $\epsilon$ again being a short distance spacetime regulator as in Eq.~\eqref{epsdef}. 
Thus the Strebel area ($A_{ij}=2L_{ij}n_{ij}$) of each such strip gives a Nambu-Goto weight 
\be\label{streb-prop}
e^{-2\pi A_{ij}} = \epsilon^{2n_{ij}} \left(\frac{1}{x_{ij}^2}\right)^{n_{ij}} \ .
\ee
As before, we absorb the regulator dependence in a multiplicative renormalisation of the corresponding YM operators. 

We therefore conclude that the worldsheets weighted with the Nambu-Goto weight in Strebel gauge give rise to the individual propagator contributions to the Feynman diagram for the Yang-Mills correlator in Eq.~\eqref{feyncorr}.  The sum over the differrent terms in  Eq.~\eqref{feyncorr} is now interpreted as a sum over different points in the moduli space which admit covering maps. These are  specified by different numbers $\{n_{ij}\}$ of wick contractions or equivalently, from the closed string point of view, the integer Strebel lengths characterising these points on moduli space.  

\section{Concluding remarks}\label{sec.6}

In this work, we have begun the task of pushing the proposal of \cite{Gaberdiel:2021qbb, Gaberdiel:2021jrv} beyond that of the agreement of the spectrum. Our goal here was to develop a geometric picture of the twistor description and see how far this can reproduce the correlators of free super Yang-Mills. This needs to be taken further and in a more systematic way at the level of the quantum correlators. Nevertheless, we already see many striking elements, through our analysis of classical configurations, which we expect will hold exactly in the full description. In developing this picture, we have also fleshed out the covering map description of $AdS_3/CFT_2$ in terms of the worldsheet twistor fields. The resulting classical twistor configurations are seen to have only the wedge modes excited as expected from the fact that these are the physical modes of the string theory after gauge fixing (as in Sec. 5 of \cite{Gaberdiel:2021jrv}). 

We then saw that this geometric picture admits a natural generalisation to ${\rm AdS}_5$. We described the natural set of incidence relations that the classical worldsheet ambitwistor configurations should satisfy. Imposing euclidean reality conditions allowed us to solve for the ${\rm AdS}_5$ spacetime string configurations in terms of the ambitwistor fields. Restricting to boundary correlators which lie on a two plane, we could write down the solutions, as in ${\rm AdS}_3$,  in terms of holomorphic covering maps with the right branching behaviour. The corresponding twistor fields are then essentially polynomials, again supporting the identification of the physical modes  with the finite number of wedge modes \cite{Gaberdiel:2021jrv, Gaberdiel:2021qbb}. We take this as a robust indication that the wedge twistor modes are indeed the crucial physical degrees of freedom in the tensionless limits of strings on both ${\rm AdS}_3$ and ${\rm AdS}_5$. It will, of course, be important to corroborate this conclusion from a first principles worldsheet analysis. 

Furthermore, for this kinematic configuration, we found that the explicit covering map is associated with the special integer points on the moduli space (`arithmetic curves'). This is because the Schwarzian of the covering map turned out to be the Strebel differential (as for $AdS_3/CFT_2$, see Fig. \ref{fig:flowchart}). The corresponding Strebel graph (or more precisely, its dual) is associated to the Feynman diagram of the free Yang-Mills theory as per \cite{Gopakumar:2005fx, Razamat:2008zr}. This realises the picture of open-closed string duality of  \cite{Gopakumar:2003ns, Gopakumar:2004qb, Gopakumar:2005fx}, associating individual 'tHooft-Feynman diagrams (open strings) with closed string worldsheets. One of the new things we learnt is that the Nambu-Goto weight (the area in Strebel gauge) associated with each such worldsheet is precisely the free Yang-Mills propagator (after proper regularisation). This is, to us, compelling evidence that the Strebel construction paves the path between perturbative field theories and closed string worldsheet theories in broad generality. 

There are a number of questions which merit further investigation, which have mostly been mentioned already. Perhaps the most important of them is to place the stringy ${\rm AdS}_5$ twistor incidence relations on a firm footing, from the standpoint of worldsheet correlators, as in \cite{Dei:2020zui}. The exact nature of the spacetime covering maps when one goes away from our special kinematic configuration is another important point to understand better. This will begin to kick in at the level of 4-point functions. As mentioned, the maps $X^{\alpha\dot{\beta}}(z, \bar{z})$ are now holomorphic in a subtle way, in there being locally a holomorphic eigenvector. Can we still use the power of holomorphy and constrain these maps? The radial profile of these configurations, which are infinitesimally near the boundary, would also be interesting to understand and should correspond to some kind of effective Liouville mode on the worldsheet. We note that the picture of the worldsheet configurations is somewhat like that described in \cite{Berkovits:2019ulm}. The connection to the Feynman diagrams for these more general configurations also needs to be worked out. Can we, for instance, see the detailed spin dependent numerators of the Feynman weights, for a general correlator, from the Strebel construction for the twistor worldsheet? It would be very nice to develop a general dictionary for the usual  Feynman rules, that we can expect to hold in any open-closed string duality in the perturbative field theory limit. 

\acknowledgments

R.G. thanks Matthias Gaberdiel for very helpful discussions on multiple matters and related collaboration. The work of R.G. is supported in part by the J.C.~Bose Fellowship of the DST-SERB.
R.G. and P.M. acknowledge the support of the Department of Atomic Energy, Government
of India, under project no.~RTI4001, as well as the framework of support for the basic sciences by the people of India. F.B. and B.R. thank ICTS-TIFR for the opportunity to be a part of the Long Term Visiting Students Program (LTVSP) that made this collaboration possible.

\appendix
\section{Conventions}\label{appA}

The spin group  $SO(4,\mathds{C})$ of complexified Minkowski space (denoted as $M_{\mathds{C}}$) is locally isomorphic to $SL(2,\mathds{C})\times SL(2,\mathds{C})$. Given a four-vector $T^{\mu}=(T^0,T^1,T^2,T^3)$, we can represent it in terms of Pauli matrices ${\sigma_{\mu}}^{\alpha\dot{\alpha}}$ as
\begin{equation}\label{vec}
    T^{\alpha\dot{\alpha}}=\frac{1}{\sqrt{2}}T^{\mu}\sigma_{\mu}^{\alpha\dot{\alpha}}=\begin{bmatrix}
            T^0+T^3 && T^1-iT^3\\
            T^1+iT^2 && T^0-T^3
    \end{bmatrix}
\end{equation}
where un-dotted and dotted spinor indices $\{\alpha=0,1|\dot{\alpha}=0,1\}$ are in the $(\frac{1}{2},0)$ and $(0,\frac{1}{2})$ representations of $SL(2,\mathds{C})\times SL(2,\mathds{C})$, known as negative and positive chirality spinors respectively. Here and below we will be largely following the conventions of \cite{Adamo:2017qyl}.

For lowering and raising the spinor indices, we can use the $SL(2,C)$-invariant tensors, i.e the usual Levi-Civita symbols 
\begin{equation}
    \epsilon_{\alpha\beta}=\epsilon^{\alpha\beta}=\epsilon_{\dot{\alpha}\dot{\beta}}=\epsilon^{\dot{\alpha}\dot{\beta}}=\begin{bmatrix}
    0 &1\\
    -1 & 0
    \end{bmatrix}.
    \end{equation}
With this convention for inverse marices $\epsilon^{\alpha\beta}$ and $\epsilon^{\dot{\alpha}\dot{\beta}}$,
\begin{equation}
    \epsilon^{\alpha\beta}\epsilon_{\gamma\beta}=\delta_{\gamma}^{\alpha} \quad \text{and}\quad \epsilon^{\alpha\beta}\epsilon_{\alpha\beta}=2
\end{equation}
and similarly for dotted indices. Our convention for lowering and raising of spinor indices will follow the slogan, `\textit{lower to the right, raise to the left}': 
\begin{equation}
    u_{\alpha}=u^{\beta}\epsilon_{\beta\alpha}\quad \text{and}\quad v^{\alpha}=\epsilon^{\alpha\beta}v_{\beta} \ ,
\end{equation}
and similarly for dotted indices. The epsilon tensors define $SL(2,\mathds{C})$ invariant inner products in the spaces of both positive and negative chirality spinors separately. We define them as
\be
\langle u\,v\rangle :=  u^{\alpha}\,v_{\alpha}=u^{\alpha}v^{\beta}\epsilon_{\beta \alpha} \quad \text{and} \quad  [\tilde{u}\,\tilde{v}]:=\tilde{u}^{\dot{\alpha}} \tilde{v}_{\dot{\alpha}}= \tilde{u}^{\dot{\alpha}}v^{\dot{\beta}}\epsilon_{\dot{\beta}\dot{\alpha}}.
\ee
We note that these inner products are skew-symmetric, i.e $\langle u\,v\rangle=-\langle v\, u \rangle$ (and similarly for the dotted spinors).
\par
The space-time vector $x^{\mu}$ in $M_{\mathds{C}}$ has the form, as in Eq. (\ref{vec})
\begin{equation}
    x^{\alpha\dot{\beta}}=\frac{1}{\sqrt{2}}\begin{bmatrix}
        x^0+x^3 && x^1-ix^2\\
        x^1+ix^2 && x^0-x^3
    \end{bmatrix} \ .
\end{equation}
Lowering and raising different indices, this becomes
\begin{equation}
    x_{\alpha}^{\;\;\;\dot{\beta}}=x^{\gamma\dot{\beta}}\epsilon_{\gamma\alpha}=\frac{1}{\sqrt{2}}\begin{bmatrix}
        -(x^1+ix^2) && -(x^0-x^3)\\
        x^0+x^3 && x^1-ix^2
    \end{bmatrix} ;
\end{equation}
\begin{equation}
    x^{\dot{\alpha}}_{\;\;\;\beta}=x^{\dot{\alpha}\gamma}\epsilon_{\gamma\beta}=\frac{1}{\sqrt{2}}\begin{bmatrix}  
        -(x^1+ix^2)&& x^0+x^3\\
        -(x^0-x^3)&& x^1-ix^2 
    \end{bmatrix} \ .
\end{equation}
This leads to the following identities
\begin{equation}\label{identity}
    \begin{split}
        & x_{\alpha\dot{\gamma}}x^{\dot{\gamma}}_{\;\;\;\beta}=\frac{1}{2}x^2 \epsilon_{\alpha\beta} \quad \text{and} \quad x^{\dot{\alpha}\gamma}x_{\gamma}^{\;\;\; \dot{\beta}}=-\frac{1}{2}x^2 \epsilon^{\dot{\alpha}\dot{\beta}} \ .
    \end{split}
\end{equation}


So far our discussion applies to $M_{\mathds{C}}$. Taking an euclidean slice of this complexified space amounts to setting
\begin{equation}\label{complex}
    x^{\alpha\dot{\beta}}=\widehat{x}^{\,\alpha\dot{\beta}},
\end{equation}
where
\begin{align}\label{hatxdef}
    \widehat{x}^{\,\alpha\dot{\beta}}=\frac{1}{\sqrt{2}}\begin{bmatrix}
           \bar{x}^0-\bar{x}^3 && -\bar{x}^1+i\bar{x}^2\\
           -\bar{x}^1-i\bar{x}^2 && \bar{x}^0+\bar{x}^3
    \end{bmatrix}.
\end{align}
This is because Eq. (\ref{complex}) implies $\bar{x}^0=x^0$, $\bar{x}^i=-x^i\, (i=1,2,3)$ so that $x^0=y^0,\, x^i=-iy^i \, (i=1,2,3)$ with $y^0,y^i\in \mathds{R}$ giving the euclidean metric $ds^2=(dy^0)^2+\sum_{i=1}^{3}(dy^i)^2$. Thus $x^{\dot{\alpha}}_{\;\;\;\beta}$ in this Euclidean real slice becomes
\begin{equation}
    x^{\dot{\alpha}}_{\;\;\;\beta}=\frac{1}{\sqrt{2}}\begin{bmatrix}
            -y^2+iy^1 && y^0-iy^3\\
            -y^0-iy^3 && -y^2-iy^1
    \end{bmatrix} \ .
\end{equation}
The stringy generalization of this expression is
\begin{equation}\label{Z-Y}
    X^{\dot{\alpha}}_{\;\;\;\beta}(z, \bar{z})=
\begin{bmatrix}
        -V(z, \bar{z}) && U(z, \bar{z}) \\
         -\bar{U}(z, \bar{z}) && -\bar{V}(z, \bar{z}) 
\end{bmatrix} \ .
\end{equation}
where we have defined $U(z,\bar{z})=(y^0-iy^3)(z,\bar{z})$ and $V(z,\bar{z})=(y^2-iy^1)(z,\bar{z})$. 

In Fig. \ref{fig:fibration_on_boundary} we have illustrated the double fibration that underlies the twistor correspondence on the Minkowski boundary of ${\rm AdS}_5$. On the left hand side is the complexified case while the RHS is after imposing the euclidean reality condition. In the latter, case the second fibration becomes a bijection between the twistor space and the spin bundle over real Euclidean space. In Fig. \ref{fig:fibration_in_bulk}, the analogous figures are shown for the complexified ${\rm AdS}_5$ spacetime and after imposing euclidean reality conditions. 

\begin{figure}[!htb]
  \begin{center}
  \begin{tikzcd}[column sep=small]
&  \underset{\textcolor{red}{(4+1)}}{\{ x^{\dot{\alpha}}_{\;\;\;\beta},\lambda^{\beta}\}} \arrow[dl, "\pi_2", rightarrow, cyan]\arrow[dr, "\pi_1", rightarrow, violet] & \\  \underset{\textcolor{red}{(3)}}{\{\mu^{\dot{\alpha}},\lambda^{\beta}\}} && \underset{ \textcolor{red}{(4)}}{x^{\dot{\alpha}}_{\;\;\;\beta}}
\end{tikzcd}
\hspace{10mm}
\begin{tikzcd}[column sep=small]
&  \underset{\textcolor{blue}{(4+2)}}{\{ x^{\dot{\alpha}}_{\;\;\;\beta},\lambda^{\beta}\}} \arrow[dl, "\pi_2", Leftrightarrow, cyan]\arrow[dr, "\pi_1", rightarrow, violet] & \\  \underset{\textcolor{blue}{(3+3)}}{\{\mu^{\dot{\alpha}},\lambda^{\beta}\}} && \underset{ \textcolor{blue}{(4)}}{x^{\dot{\alpha}}_{\;\;\;\beta}}
\end{tikzcd}
\end{center}
    \caption{(Left) Double fibration of the projective spinor  bundle $\{ x^{\dot{\alpha}}_{\;\;\;\beta},\lambda^{\beta}\}\cong M_{\mathds{C}}\times \mathds{CP}^1$ over the twistor space $\{\mu^{\dot{\alpha}},\lambda^{\beta}\}\cong \mathds{CP}^3 $ and the space-time on the boundary of $AdS_5$, $\{x^{\dot{\alpha}}_{\;\;\;\beta}\}\cong M_{\mathds{C}}$. Numbers in red are the complex dimensions of the corresponding spaces. (Right) The same fibration after imposing the euclidean reality conditions.  Numbers in blue now show the real dimensions of the corresponding spaces. Note that $\pi_2$ is an isomorphism in this signature.}
    \label{fig:fibration_on_boundary}
\end{figure}
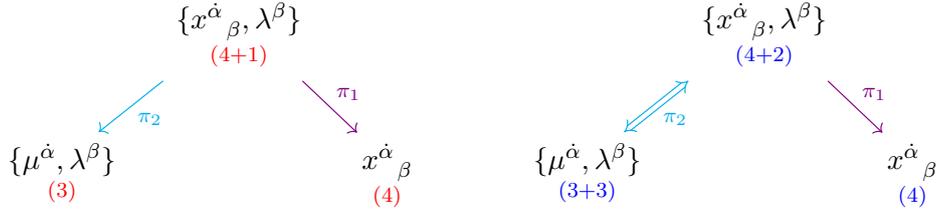

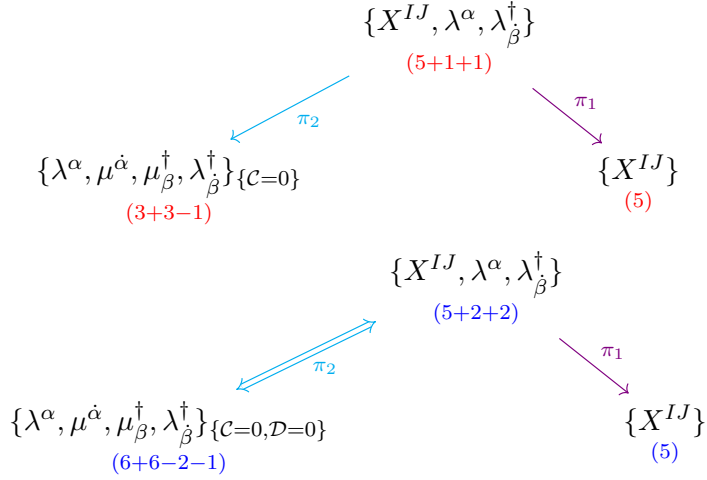
\begin{figure}[!htb]
  \begin{center}
  \begin{tikzcd}[column sep=small]
&  \underset{\textcolor{red}{(5+1+1)}}{\{ X^{IJ},\lambda^{\alpha},\lambda_{\dot{\beta}}^{\dagger}\}} \arrow[dl, "\pi_2", rightarrow,  cyan]\arrow[dr, "\pi_1", rightarrow, violet] & \\  \underset{\textcolor{red}{(3+3-1)}}{\{\lambda^{\alpha},\mu^{\dot{\alpha}},\mu^{\dagger}_{\beta},\lambda^{\dagger}_{\dot{\beta}}\}_{\{\mathcal{C}=0\}}} && \underset{\textcolor{red}{(5)}}{\{ X^{IJ}\}}
\end{tikzcd}
\vspace{2mm}
\begin{tikzcd}[column sep=small]
&  \underset{\textcolor{blue}{(5+2+2)}}{\{ X^{IJ},\lambda^{\alpha},\lambda_{\dot{\beta}}^{\dagger}\}} \arrow[dl, "\pi_2", Leftrightarrow, cyan]\arrow[dr, "\pi_1", rightarrow, violet] & \\  \underset{\textcolor{blue}{(6+6-2-1)}}{\{\lambda^{\alpha},\mu^{\dot{\alpha}},\mu^{\dagger}_{\beta},\lambda^{\dagger}_{\dot{\beta}}\}_{\{\mathcal{C}=0, \mathcal{D}=0\}}} && \underset{\textcolor{blue}{(5)}}{\{ X^{IJ}\}}
\end{tikzcd}
\end{center}
    \caption{(Top) Double fibration of the projective spinor bundle $\{ X^{IJ},\lambda^{\alpha},\lambda_{\dot{\beta}}^{\dagger}\}\cong \mathds{CP}^5\times \mathds{CP}^1\times \mathds{CP}^1$ over the Ambi-twistor space $\{\lambda^{\alpha},\mu^{\dot{\alpha}},\mu^{\dagger}_{\beta},\lambda^{\dagger}_{\dot{\beta}}\}_{\{\mathcal{C}=0\}}\cong (\mathds{CP}^3\times \mathds{CP}^3)_{\{\mathcal{C}=0\}}$ and the ``space-time" $\{X^{IJ}\}\cong \mathds{CP}^5$ in the complexified $AdS_5$ bulk. ${\cal C}=0$ refers to the ambitwistor constraint Eq. \eqref{bulkC}. Numbers in red are the complex dimensions of the corresponding spaces. (Bottom)  The same fibration after imposing the euclidean reality conditions which leads to the additional constraint ${\cal D}=0$ -- see Eq. \eqref{bulk-constr}.  Numbers in blue are the real dimensions of the corresponding spaces. Note that $\pi_2$ is an isomorphism in this signature.}
    \label{fig:fibration_in_bulk}
\end{figure}

\section{Feynman covering maps in the symmetric orbifold $CFT_2$}\label{appB}

Here we give a brief recap of the computation of the correlators in symmetric orbifold theories from covering maps \cite{Lunin:2001} and their associated ``Feynman graphs"\cite{Pakman:2009}. We then summarise the results of \cite{Gaberdiel:2020ycd} where it was shown that each such Feynman covering map (at least, for large twists) can be associated to a point in the closed string moduli space via the Strebel correspondence, thus realising the open-closed string dictionary of \cite{Gopakumar:2005fx, Razamat:2008zr}. In the present case, these points are precisely the ones which admit covering maps, and to which the closed string correlator localises \cite{Eberhardt:2019ywk, Dei:2020zui}. A new geometric ingredient in the correspondence, is that the Strebel differential, in the large twist limit, is nothing other than the Schwarzian of the covering map. 

\subsection{Symmetric orbifold CFT correlators and branched coverings}

We consider the free symmetric orbifold CFT ${\rm Sym}^K(\mathbb{\mathcal{S}})$ in the large $K$ limit. The orbifold theory consists of states from both twisted and untwisted sectors where single cycle twist fields correspond to the single string states in the bulk $AdS_3$. We will mainly be interested in the correlators of the ground states of such single cycle twist operators $\{\sigma_{w_i}\}$ 
\begin{equation}\label{correlator1}
\langle \sigma_{w_1}(x_1)\cdots \sigma_{w_n}(x_n)\rangle .
\end{equation}
Here $x_i$ are coordinates on the space-time $\mathbb{S}^2$ and the subscripts label the cycle lengths. 

A $w$-cycle twist field $\sigma_w(x_0)$ induces a $w$-fold cyclic permutation amongst some of the $K$ copies of the $\mathcal{S}$ theory. The action of this permutation can be geometrised by locally lifting to a $w$-fold covering near any of the insertion points. Globally, these can be combined into a finite degree covering surface $\Sigma$ described by the covering map 
 \begin{equation}
    \Gamma:\Sigma\,[z]\to \mathbb{S}^2\,[x]\, .
\end{equation}
The covering map has the local branching behaviour in the vicinity of a pre-image $z_i$ of an insertion point $x_i$
 \begin{equation}
    \Gamma(z) \approx x_i +(z-z_i)^{w_i}.
\end{equation}
Restricting to genus zero covering surfaces (which corresponds to the leading order in a $1/K$ expansion of the correlator) fixes the degree $N$, i.e total number of sheets in the lift of the correlator,  by the Riemann-Hurwitz formula 
\begin{equation}
    N=1+\frac{1}{2}\sum_{j=1}^{n}(w_j-1)\, .
\end{equation}

The prescription put forward by \cite{Lunin:2001} is to exploit this covering map to uplift the calculation of the correlator to the covering surface. On the covering surface the ground state twist fields $\sigma_{w_i}$ become identity operators, since their associated monodromy is captured by the covering surface itself. Thus the correlator on the covering surface simply becomes the vacuum path integral getting contributions only from the Liouville anomaly term for the conformal factor $\partial \Gamma(z)$, associated to the covering map.
    \begin{equation}\label{correlator}
        \langle  \sigma_{w_1}(x_1) \cdots  \sigma_{w_n}(x_n) \rangle = \sum_\Gamma W_{\Gamma}\, e^{-S_{\rm L}[{\Phi}_\Gamma]} \ , 
    \end{equation}
    where $S_{L}[\Phi]$ is the familiar Liouville 
    action
    \begin{equation}
        S_{\rm L}[\Phi] = \frac{c}{48 \pi} \int d^2z \sqrt{g} \bigl( \, 2 \, \partial \Phi \, \bar\partial \Phi + R \, \Phi \Bigr) \,
    \end{equation}
    and $${{\Phi}}_\Gamma = \log \, \partial_z \Gamma(z)  + \log \, \partial_{\bar{z}} \bar{\Gamma}(\bar{z})\, .$$
The sum over $\Gamma$ in Eq. \eqref{correlator} is over the finitely many inequivalent covering maps that exist for specified branching data, namely, $\{w_i, x_i \}$. Specifying these fixes $(n-3)$ locations $z_i$ on the covering space (three points can be fixed using the global Mobius invariance). In other words, we get contributions from only discretely many points on the moduli space of the covering space. In \cite{Eberhardt:2019ywk}, \cite{Dei:2020zui}, these points were seen to be precisely the ones which give a non-zero contribution to the worldsheet correlator. This cements the identification of the covering space with the worldsheet as conjectured already in \cite{Lunin:2001, Pakman:2009}.

\subsection{Branched coverings and Feynman diagrams for the orbifold $CFT_2$}

Next, we recollect the diagrammatic picture of orbifold correlators introduced in \cite{Pakman:2009}. The basic idea is to consider a double-lined (dashed and solid) Jordan curve on the base sphere $\mathbb{S}^2[x]$ passing through the $n$ insertion points $x_i$ with the coloured (solid border) face enclosing $x=\infty$ (see the right hand figure in Fig. \ref{fig:Orbifold_feynmans}). Then the pre-image of this curve under the covering map $\Gamma$ defines a `Feynman graph' for the contribution from this particular $\Gamma$ to the correlator (\ref{correlator}). This graph has bi-fundamental lines (edges) joining the different vertices (located at the $z_i$). The poles of $\Gamma$ being pre-images of $x=\infty$ are now located in the interior of the $N$ colored faces in the Feynman graph. Thus these bifundamental Feynman graphs give a bi-colouration of the worldsheet Riemann surface. 

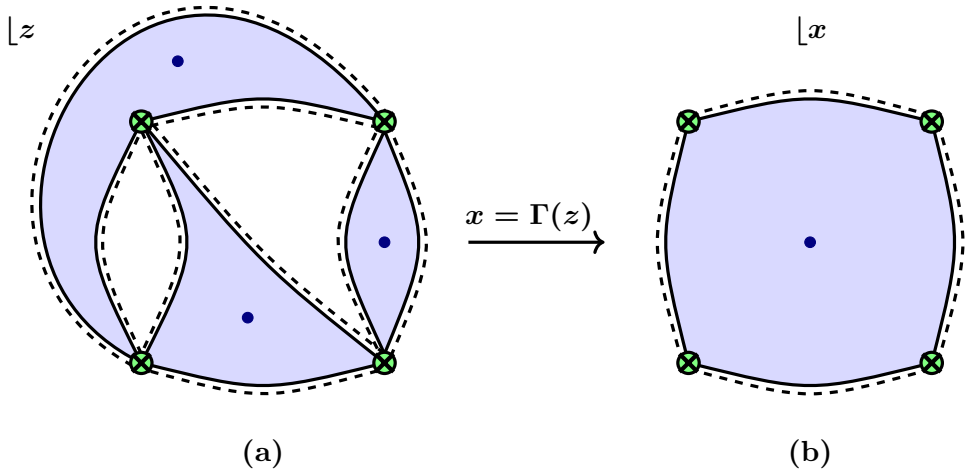
\begin{figure}[htb]
\vspace*{-2.5cm}
\hspace*{-2.1cm}
\centering
\begin{tikzpicture}[scale = 0.8]



\draw[black, very thick, fill=blue!15!] (4,4)..controls(0,9) and (-4,2)..(0,0)..controls(-1,2)..(0,4)..controls (2,4.5)..(4,4);

\draw[black, very thick, fill=blue!15!] (0,0)..controls (2,-0.5)..(4,0)..controls (2,1.6)..(0,4)..controls (1,2)..(0,0);

\draw[black, very thick, fill=blue!15!] (4,0.15)..controls (3.15,2)..(4,3.85)..controls (4.75,2)..(4,0.15);


\draw[black, very thick, dashed] (0,0.15)..controls (-0.85,2)..(0,3.85)..controls (0.85,2)..(0,0.15);

\draw[black, very thick, dashed] (4,0)..controls (3,2)..(3.9,3.9);

\draw[black, very thick, dashed] (4,0.15)..controls (2,1.75)..(0.15,4);

\draw[black, very thick, dashed] (0.15,3.93)..controls (2,4.35)..(3.9,3.9);

\draw[black, very thick, dashed] (-0.1,-0.07)..controls (2,-0.65)..(4,-0.1);

\draw[black, very thick, dashed] (4.1,0.1)..controls (4.9,2)..(4,4.1);

\draw[black, very thick, dashed] (4.07,4.1)..controls (0,9.15) and (-4.25,2.15)..(-0.1,-0.07);

\filldraw [color=black, fill=green!50, very thick] (4,4) circle (5pt) node[cross,black] {};
\filldraw [color=black, fill=green!50, very thick] (0,0) circle (5pt) node[cross,black] {};
\filldraw [color=black, fill=green!50, very thick]  (0,4) circle (5pt) node[cross,black] {};
\filldraw [color=black, fill=green!50, very thick]  (4,0) circle (5pt) node[cross,black] {};

\filldraw[blue!50!black!100!] (4,2) circle (2.5pt);
\filldraw[blue!50!black!100!] (1.75,0.75) circle (2.5pt);
\filldraw[blue!50!black!100!] (0.6,5) circle (2.5pt);

\draw[black, very thick,->] (5.38,2)--(7.62,2) node[anchor=south east] {$\boldsymbol{x=\Gamma(z)}$};


\draw[black, very thick, fill=blue!15!] (9,0)..controls (11,-0.5)..(13,0)..controls (13.5,2)..(13,4)..controls (11,4.5)..(9,4)..controls (8.5,2)..(9,0);

\draw[black, very thick, dashed] (9,-0.1)..controls (11,-0.65)..(13,-0.1);

\draw[black, very thick, dashed] (13.1,0)..controls (13.65,2)..(13.1,4);

\draw[black, very thick, dashed] (13,4.1)..controls (11,4.65)..(9,4.1);

\draw[black, very thick, dashed] (8.9,4)..controls (8.35,2)..(8.9,0);

\filldraw [color=black, fill=green!50, very thick] (13,4) circle (5pt) node[cross,black] {};
\filldraw [color=black, fill=green!50, very thick] (9,0) circle (5pt) node[cross,black] {};
\filldraw [color=black, fill=green!50, very thick] (13,0) circle (5pt) node[cross,black] {};
\filldraw [color=black, fill=green!50, very thick] (9,4) circle (5pt) node[cross,black] {};

\filldraw[blue!50!black!100!] (11,2) circle (2.5pt);

\node[] at (-2, 5.5)  {$\boldsymbol{\lfloor z}$};
\node[] at (11, 5.5)  {$\boldsymbol{\lfloor x}$};

\node[] at (2, -1.5) {\textbf{(a)}};
\node[] at (11, -1.5) {\textbf{(b)}};
\end{tikzpicture}
\caption{Feynman graph of a symmetric orbifold correlator, formed by the preimage of a Jordan curve under $\Gamma$. Here $n=4$, $w_i=2$ and so $N=3$. Figure taken from \cite{Gaberdiel:2020ycd}}.
\label{fig:Orbifold_feynmans}
\end{figure}

The sum over the inequivalent covering maps is nothing other than a sum over the  inequivalent Feynman graphs, subject to the diagrammatic constraints explained in \cite{Pakman:2009}. We have a one to one correspondence of each Feynman graph with a particular worldsheet (point in the moduli space which admits the corresponding covering map). We will now see that this association and the geometric picture of gluing up of these Feynman graphs to form the worldsheet is captured, for large twist $w_i$, by the Strebel construction as in \cite{Gopakumar:2005fx, Razamat:2008zr}.  

\subsection{The large twist limit and the Strebel construction}

While the Lunin-Mathur approach to the orbifold CFT correlators is powerful in principle, it is difficult in  practice since one needs to explicitly obtain the covering map $\Gamma(z)$. In \cite{Gaberdiel:2020ycd} it was shown that this problem simplifies in the large $N$ i.e. large twist limit. Let us recap the main steps in \cite{Gaberdiel:2020ycd}. For an $n$-point function (\ref{correlator1}) (with $z_n=\Gamma^{-1}(x_n)=\infty$), we have   
\begin{equation}\label{partG}
\partial \Gamma(z) = M_{\Gamma} \, \frac{ \prod_{i=1}^{n-1}(z-z_i)^{w_i-1}}{\prod_{a=1}^{N}(z-\lambda_a)^{2}} \ ,
\end{equation}
where $M_{\Gamma}$ is a non-zero constant. Following \cite{Roumpedakis:2018tdb}, the condition that the $\lambda$ are simple poles of $\Gamma(z)$ leads to $N$ ``scattering equations" 
\begin{equation}\label{scatt}
\sum_{i=1}^{n-1} \frac{w_i-1}{\lambda_a-z_i}=\sum_{b\neq a}^{N} \frac{2}{\lambda_a-\lambda_b} \ , \qquad  (a=1,\ldots, N) \ .
\end{equation}
In principle, these determine the poles $\{\lambda_a\}$ in (\ref{partG}) and therefore the covering map (upto overall additive and multiplicative factors). The crucial observation of \cite{Gaberdiel:2020ycd} was that, in the limit of large $w_i$, these are the saddle-point equations of a Penner-like matrix model, with a logarithmic potential. We can then compute the covering map in terms of the `spectral curve' $y_0(z)$ of the matrix model. To leading order in $\frac{1}{N}$ 
\begin{equation}
    \frac{1}{N}\partial \ln \partial \Gamma = y_0(z)\ ,
\end{equation}
The spectral curve $y_0(z)$ is a hyperelliptic curve with cuts formed from the coalesced poles $\{\lambda_a\}$ of $\Gamma(z)$ in the large $N$ limit. This curve is therefore specified by $2(n-3)$ numbers which are the A and B-cycle periods around the cuts. These parameters serve as the free moduli characterizing the different covering maps that contribute to the original correlator. 

Diagrammatically, the situation is as shown in fig.\,\ref{fig:twofigures}. The coalescing poles form transversal cuts across the edges of the original Wick contractions between vertices of the Feynman graph (see the right hand figure in \ref{fig:twofigures}). Thus the cuts of the spectral curve correspond to the edges of the dual to the original (Skeleton) Feynman graph. The A- and B-cycle periods of the spectral curve simply count the number of eigenvalues i.e. poles. This number, for large twist, is simply given by the number of Wick contractions of the edges (see left figure in \ref{fig:twofigures}) :
\begin{equation}\label{periodintegral}
\oint_{C_{ij}} \frac{dz}{2\pi i}\, y_0(z)=\frac{n_{ij}}{N}\ ,
\end{equation}
where $n_{ij}$ is the number  of Wick contractions in the edge connecting $i$-th and $j$-th vertex.
Note that in the strict large $N$ limit, these periods can take arbitrary {\it real} values (upto an overall scaling). 

\begin{figure}[!htb]
    \centering
    \begin{subfigure}{0.55\textwidth}
    \hspace{-20mm}
  \begin{tikzpicture}[scale = 0.7]
\draw[very thick, fill=blue!15!] (0,0) to[out = 20, in = -20,  looseness=2.5] (0,6) to[out=-30, in=30, looseness=2.2] (0,0);

\draw[very thick, dashed] (0,0-0.05) to[out = 20-3, in = -20+3,  looseness=2.5] (0,6+0.05) (0,6-0.05) to[out=-30-3, in=30+3, looseness=2.2] (0,0+0.05);

\draw[very thick, fill=blue!15!] (0,0) to[out = 160, in = -160,  looseness=2.5] (0,6) to[out=-150, in=150,looseness=2.2] (0,0);

\draw[very thick, dashed] (0,0-0.05) to[out = 160+3, in = -160-3,  looseness=2.5] (0,6+0.05) (0,6-0.05) to[out=-150+3, in=150-3, looseness=2.2] (0,0+0.05);


\draw[very thick, fill=blue!15!] (0,0) to[out = 40, in = -40,  looseness=1.9] (0,6) to[out=-45, in=45, looseness=1.6] (0,0);

\draw[very thick, dashed] (0,0-0.05) to[out = 40-3, in = -40+3,  looseness=1.9] (0,6+0.05) (0,6-0.05) to[out=-45-3, in=45+3, looseness=1.6] (0,0+0.05);

\draw[very thick, fill=blue!15!] (0,0) to[out = 180-40, in = -180+40,  looseness=1.9] (0,6) to[out=-180+45, in=180-45, looseness=1.6] (0,0);

\draw[very thick, dashed] (0,0-0.05) to[out = 180-40+3, in = -180+40-3,  looseness=1.9] (0,6+0.05) (0,6-0.05) to[out=-180+45+3, in=180-45-3, looseness=1.6] (0,0+0.05);

\draw[very thick, fill=blue!15!] (0,0) to[out = 55, in = -55,  looseness=1.3] (0,6) to[out=-58, in=58, looseness=1] (0,0);

\draw[very thick, dashed] (0,0-0.05) to[out = 55-3, in = -55+3,  looseness=1.3] (0,6+0.05) (0,6-0.05) to[out=-58-3, in=58+3, looseness=0.95] (0,0+0.05);

\draw[very thick, fill=blue!15!] (0,0) to[out = 180-55, in = -180+55,  looseness=1.3] (0,6) to[out=-180+58, in=180-58, looseness=1] (0,0);

\draw[very thick, dashed] (0,0-0.05) to[out = 180-55+3, in = -180+55-3,  looseness=1.3] (0,6+0.05) (0,6-0.05) to[out=-180+58+3, in=180-58-3, looseness=0.95] (0,0+0.05);

\draw[very thick, fill=blue!15!] (0,0) to[out = 80, in = -80,  looseness=0.8] (0,6) to[out=-180+80, in=180-80, looseness=0.8] (0,0);

\draw[very thick, dashed] (0+0.05,0+0.05) to[out = 80-3, in = -80+3,  looseness=0.85] (0+0.05,6) (0-0.05,6) to[out=-180+80-3, in=180-80+3, looseness=0.85] (0-0.05,0);


\draw[color=white, fill=blue!15!] (0,6)..controls (-0.8,6.8)..(-2,6.8)--(-1.5,7.5)..controls (-0.5,7)..(0,6);
\draw[very thick] (0,6)..controls (-0.8,6.8)..(-2,6.8) (-1.5,7.5)..controls (-0.5,7)..(0,6);
\draw[very thick, dashed] (-0.1,6)..controls (-0.8,6.7)..(-2,6.7) (-1.5,7.6)..controls (-0.5,7.1)..(0.1,6);

\draw[color=white, fill=blue!15!] (0,6)..controls (0.8,6.8)..(2,6.8)--(1.5,7.5)..controls (0.5,7)..(0,6);
\draw[very thick] (0,6)..controls (0.8,6.8)..(2,6.8) (1.5,7.5)..controls (0.5,7)..(0,6);
\draw[very thick, dashed] (0.1,6)..controls (0.8,6.7)..(2,6.7) (1.5,7.6)..controls (0.5,7.1)..(-0.1,6);

\draw[color=white, fill=blue!15!] (-0.02,6)..controls (-0.1,7.3)..(-0.3,8)--(0.3,8)..controls (0.1,7.3)..(0.1,6)--(0.02,6);
\draw[very thick] (-0.02,6)..controls (-0.1,7.5)..(-0.3,8) (0.3,8)..controls (0.1,7.5)..(0.02,6)--(0,6);
\draw[very thick, dashed] (-0.1,6)..controls (-0.2,7.5)..(-0.4,8) (0.4,8)..controls (0.2,7.5)..(0.1,6);


\draw[color=white, fill=blue!15!] (0,6-6)..controls (-0.8,-6.8+6)..(-2,-6.8+6)--(-1.5,-7.5+6)..controls (-0.5,-7+6)..(0,-6+6);
\draw[very thick] (0,6-6)..controls (-0.8,-6.8+6)..(-2,-6.8+6) (-1.5,-7.5+6)..controls (-0.5,-7+6)..(0,-6+6);
\draw[very thick, dashed] (-0.1,-6+6)..controls (-0.8,-6.7+6)..(-2,-6.7+6) (-1.5,-7.6+6)..controls (-0.5,-7.1+6)..(0.1,-6+6);

\draw[color=white, fill=blue!15!] (0,-6+6)..controls (0.8,-6.8+6)..(2,-6.8+6)--(1.5,-7.5+6)..controls (0.5,-7+6)..(0,-6+6);
\draw[very thick] (0,-6+6)..controls (0.8,-6.8+6)..(2,-6.8+6) (1.5,-7.5+6)..controls (0.5,-7+6)..(0,-6+6);
\draw[very thick, dashed] (0.1,-6+6)..controls (0.8,-6.7+6)..(2,-6.7+6) (1.5,-7.6+6)..controls (0.5,-7.1+6)..(-0.1,-6+6);

\draw[color=white, fill=blue!15!] (-0.02,-6+6)..controls (-0.1,-7.3+6)..(-0.3,-8+6)--(0.3,-8+6)..controls (0.1,-7.3+6)..(0.1,-6+6)--(0.02,-6+6);
\draw[very thick] (-0.02,-6+6)..controls (-0.1,-7.5+6)..(-0.3,-8+6) (0.3,-8+6)..controls (0.1,-7.5+6)..(0.02,-6+6)--(0,-6+6);
\draw[very thick, dashed] (-0.1,-6+6)..controls (-0.2,-7.5+6)..(-0.4,-8+6) (0.4,-8+6)..controls (0.2,-7.5+6)..(0.1,-6+6);

\draw[very thick] (-6.5,3)--(-3.7,3)--(-2.95,3.5)--(-2.25,3)--(-1.68,3.5)--(-1.12,3)--(-0.56,3.5)--(0,3)--(0.56,3.5)--(1.12,3)--(1.68,3.5)--(2.25,3)--(2.95,3.5)--(3.7,3)--(6.5,3);

\filldraw[blue!50!black!100!] (0,3) circle (2.2pt) (1.12,3) circle (2.2pt)   (2.25,3) circle (2.2pt) (3.7,3) circle (2.2pt) (-1.12,3) circle (2.2pt) (-2.25,3) circle (2.2pt) (-3.7,3) circle (2.2pt);

\draw[thick, violet] (0,3) ellipse (5cm and 1.2cm);

\filldraw [color=black, fill=green!50, very thick] (0,0) circle (6pt) node[cross,black] {};
\filldraw [color=black, fill=green!50, very thick] (0,6) circle (6pt) node[cross,black] {};

\node at (1.9, -0.1)  {$\boldsymbol{j^{\text{\textbf{th}}}}$ \textbf{vertex}};
\node at (1.9, 6.2)  {$\boldsymbol{i^{\text{\textbf{th}}}}$ \textbf{vertex}};
\node at (6.6, +4.2)  {\scalebox{0.8}{$\boldsymbol{\displaystyle\oint_{C_{ij}} \frac{dz}{2\pi i}\, y_0(z)=\frac{n_{ij}}{N}}$}};

\draw[color=white, fill=white] (-0.09, 2.55) circle (0.35);
\node at (-0.09, 2.55)  {\scalebox{1.1}{$\boldsymbol{C_{ij}}$}};

\draw[->, thick, violet] (4.6,3.62)--(4.3,3.75);

\end{tikzpicture}
\end{subfigure}%
\begin{subfigure}{0.55\textwidth}
 \begin{tikzpicture}[scale = 0.55]


\fill[color = blue!15!] (-4,-3 + 0.10) to[out = 7, in = 180 - 7] (4,-3 + 0.10) --  (4 - 0.15,-3) to[out = 90 + 11, in = -90 - 11] (4 - 0.15,3) -- (4 + 0.10/1.44,3 - 0.10/1.44) to[out = 180 + 37 + 7, in = 37 - 7] (-4 + 0.10/1.44,-3 - 0.10/1.44);
\fill[color = blue!15!] (-4,3 + 0.10) to[out = 7, in = 180 - 7] (4,3 + 0.10) -- (4 + 0.15,3) to[in = 90 - 11, out = -90 + 11] (4 + 0.15,-3) -- (4 - 0.1,-3 + 0.1) to[in = 45, out = 45, distance = 8.4cm] (-4 + 0.1,3 - 0.1);
\fill[color = blue!15!] (-4,-3) to[out = -2, in = -180 + 2] (4,-3) -- (4,-3 - 0.05) to[in = -5, out = -180 + 5] (-4,-3 - 0.05);
\fill[color = blue!15!] (-4,3) to[out = -2, in = -180 + 2] (4,3) -- (4,3 - 0.05) to[in = -5, out = -180 + 5] (-4,3 - 0.05);
\fill[color = blue!15!] (4,-3) to[out = 90 - 4, in = -90 + 4] (4,3) -- (4,3) to[in = 90 + 4, out = -90 - 4] (4,-3);
\fill[color = blue!15!] (-4 + 0.05,-3) to[out = 90 - 5, in = -90 + 5] (-4 + 0.05,3) -- (-4 + 0.10,3) to[in = 90 - 10, out = -90 + 10] (-4 + 0.10,-3);
\fill[color = blue!15!] (-4 - 0.05,-3) to[out = 90 + 5, in = -90 - 5] (-4 - 0.05,3) -- (-4 - 0.10,3) to[in = 90 + 10, out = -90 - 10] (-4 - 0.10,-3);
\fill[color = blue!15!] (-4,-3) to[out = 37 + 2, in = 180 + 37 - 2] (4,3) -- (4 - 0.05/1.44,3 + 0.05/1.44) to[in = 37 + 5, out = 180 + 37 - 5] (-4 - 0.05/1.44,-3 + 0.05/1.44);
\fill[color = blue!15!] (-4 - 0.02,3 + 0.02) to[out = 45, in = 45, distance = 8.95cm] (4 + 0.02,-3 - 0.02) -- (4 + 0.05,-3 - 0.05) to[in = 45, out = 45, distance = 9.15cm] (-4 - 0.05,3 + 0.05);


\draw[very thick] (1.5,-3.6) -- (1.5,-1);
\draw[very thick] (-1.5,1) -- (-1.5,3.6);
\draw[very thick] (1.5,-1) -- (4.3,-1);
\draw[very thick] (-4.65,1) -- (-1.5,1);
\draw[very thick] (-0.75,0.75) -- (0.75,-0.75);
\draw[very thick] (-0.75,0.75) to[out = 135, in = 0] (-1.5,1);
\draw (-1.5,1) node[cross=4] {};
\draw[very thick] (0.75,-0.75) to[out = -45, in = 180] (1.5,-1);
\draw (1.5,-1) node[cross=4] {};
\draw[very thick] (4.3,-1) to[out = 0, in = -90] (4.75,3.65);
\draw[very thick] (-1.5,3.3) to[out = 90, in = 180] (4.75,3.65);
\draw (4.75,3.65) node[cross=4] {};
\draw[very thick] (4.75,3.65) -- (6,4.5);
\draw[very thick] (-4.65,1) to[out = 180, in = 190] (0,7) to[out = 10, in = 90] (6,4.5);
\draw[very thick] (1.5,-3.6) to[out = -90, in = -100] (8,0) to[out = 80, in = 0] (6,4.5);
\draw (6,4.5) node[cross=4] {};


\draw[very thick, black] (-4,-3) to[out = -2, in = -180 + 2] (4,-3);
\draw[very thick, black, dashed] (-4,-3) to[out = 2, in = 180 - 2] (4,-3);
\draw[very thick, black] (-4,-3 - 0.05) to[out = -5, in = -180 + 5] (4,-3 - 0.05);
\draw[very thick, black, dashed] (-4,-3 + 0.05) to[out = 5, in = 180 - 5] (4,-3 + 0.05);
\draw[very thick, black, dashed] (-4,-3 - 0.10) to[out = -7, in = -180 + 7] (4,-3 - 0.10);
\draw[very thick, black] (-4,-3 + 0.10) to[out = 7, in = 180 - 7] (4,-3 + 0.10);
\fill (1.5,-2.4) circle (0.07);
\fill (1.5,-3.15) circle (0.07);


\draw[very thick, black] (-4,3) to[out = -2, in = -180 + 2] (4,3);
\draw[very thick, black, dashed] (-4,3) to[out = 2, in = 180 - 2] (4,3);
\draw[very thick, black] (-4,3 - 0.05) to[out = -5, in = -180 + 5] (4,3 - 0.05);
\draw[very thick, black, dashed] (-4,3 + 0.05) to[out = 5, in = 180 - 5] (4,3 + 0.05);
\draw[very thick, black, dashed] (-4,3 - 0.10) to[out = -7, in = -180 + 7] (4,3 - 0.10);
\draw[very thick, black] (-4,3 + 0.10) to[out = 7, in = 180 - 7] (4,3 + 0.10);
\fill (-1.5,2.85) circle (0.07);
\fill (-1.5,3.5) circle (0.07);


\draw[very thick, black] (4,-3) to[out = 90 - 4, in = -90 + 4] (4,3);
\draw[very thick, black] (4,-3) to[out = 90 + 4, in = -90 - 4] (4,3);
\draw[very thick, black, dashed] (4 + 0.05,-3) to[out = 90 - 5, in = -90 + 5] (4 + 0.05,3);
\draw[very thick, black, dashed] (4 - 0.05,-3) to[out = 90 + 5, in = -90 - 5] (4 - 0.05,3);
\draw[very thick, black, dashed] (4 + 0.10,-3) to[out = 90 - 10, in = -90 + 10] (4 + 0.10,3);
\draw[very thick, black, dashed] (4 - 0.10,-3) to[out = 90 + 10, in = -90 - 10] (4 - 0.10,3);
\draw[very thick, black] (4 + 0.15,-3) to[out = 90 - 11, in = -90 + 11] (4 + 0.15,3);
\draw[very thick, black] (4 - 0.15,-3) to[out = 90 + 11, in = -90 - 11] (4 - 0.15,3);
\fill (4,-1) circle (0.07);


\draw[very thick, black, dashed] (-4,-3) to[out = 90 - 4, in = -90 + 4] (-4,3);
\draw[very thick, black, dashed] (-4,-3) to[out = 90 + 4, in = -90 - 4] (-4,3);
\draw[very thick, black] (-4 + 0.05,-3) to[out = 90 - 5, in = -90 + 5] (-4 + 0.05,3);
\draw[very thick, black] (-4 - 0.05,-3) to[out = 90 + 5, in = -90 - 5] (-4 - 0.05,3);
\draw[very thick, black] (-4 + 0.10,-3) to[out = 90 - 10, in = -90 + 10] (-4 + 0.10,3);
\draw[very thick, black] (-4 - 0.10,-3) to[out = 90 + 10, in = -90 - 10] (-4 - 0.10,3);
\draw[very thick, black, dashed] (-4 + 0.15,-3) to[out = 90 - 11, in = -90 + 11] (-4 + 0.15,3);
\draw[very thick, black, dashed] (-4 - 0.15,-3) to[out = 90 + 11, in = -90 - 11] (-4 - 0.15,3);
\fill (-4.27,1) circle (0.07);
\fill (-3.73,1) circle (0.07);


\draw[very thick, black] (-4,-3) to[out = 37 + 2, in = 180 + 37 - 2] (4,3);
\draw[very thick, black, dashed] (-4,-3) to[out = 37 - 2, in = 180 + 37 + 2] (4,3);
\draw[very thick, black] (-4 - 0.05/1.44,-3 + 0.05/1.44) to[out = 37 + 5, in = 180 + 37 - 5] (4 - 0.05/1.44,3 + 0.05/1.44);
\draw[very thick, black, dashed] (-4 + 0.05/1.44,-3 - 0.05/1.44) to[out = 37 - 5, in = 180 + 37 + 5] (4 + 0.05/1.44,3 - 0.05/1.44);
\draw[very thick, black, dashed] (-4 - 0.10/1.44,-3 + 0.10/1.44) to[out = 37 + 7, in = 180 + 37 - 7] (4 - 0.10/1.44,3 + 0.10/1.44);
\draw[very thick, black] (-4 + 0.10/1.44,-3 - 0.10/1.44) to[out = 37 - 7, in = 180 + 37 + 7] (4 + 0.10/1.44,3 - 0.10/1.44);
\fill (-0.14,0.14) circle (0.07);


\draw[very thick, black] (-4 - 0.02,3 + 0.02) to[out = 45, in = 45, distance = 8.95cm] (4 + 0.02,-3 - 0.02);
\draw[very thick, black, dashed] (-4,3) to[out = 45, in = 45, distance = 8.8cm] (4,-3);
\draw[very thick, black] (-4 - 0.05,3 + 0.05) to[out = 45, in = 45, distance = 9.15cm] (4 + 0.05,-3 - 0.05);
\draw[very thick, black, dashed] (-4 + 0.05,3 - 0.05) to[out = 45, in = 45, distance = 8.55cm] (4 - 0.05,-3 + 0.05);
\draw[very thick, black, dashed] (-4 - 0.1,3 + 0.1) to[out = 45, in = 45, distance = 9.3cm] (4 + 0.1,-3 - 0.1);
\draw[very thick, black] (-4 + 0.1,3 - 0.1) to[out = 45, in = 45, distance = 8.4cm] (4 - 0.1,-3 + 0.1);
\fill (5.48,4.15) circle (0.07);


\foreach \x in {0,1}
\foreach \y in {0,1}
{
\draw[very thick, fill = green!50] (-4 + 8*\x, -3 + 6*\y) circle (0.35);
\draw[ultra thick] (-4 + 8*\x + 0.35*0.18/0.25, -3 + 6*\y + 0.35*0.18/0.25) -- (-4 + 8*\x - 0.35*0.18/0.25, -3 + 6*\y - 0.35*0.18/0.25);
\draw[ultra thick] (-4 + 8*\x + 0.35*0.18/0.25, -3 + 6*\y - 0.35*0.18/0.25) -- (-4 + 8*\x - 0.35*0.18/0.25, -3 + 6*\y + 0.35*0.18/0.25);
}
\end{tikzpicture}
\end{subfigure}
    \caption{(Left) The poles of the covering map coalesce to form cuts across the edges of the Feynman graph where the period integral counts the number of Wick contractions. (Right)  Feynman graph and it's dual for the four-point function of each operator with twist two. Figure taken from \cite{Gaberdiel:2020ycd}.}
    \label{fig:twofigures}
\end{figure}
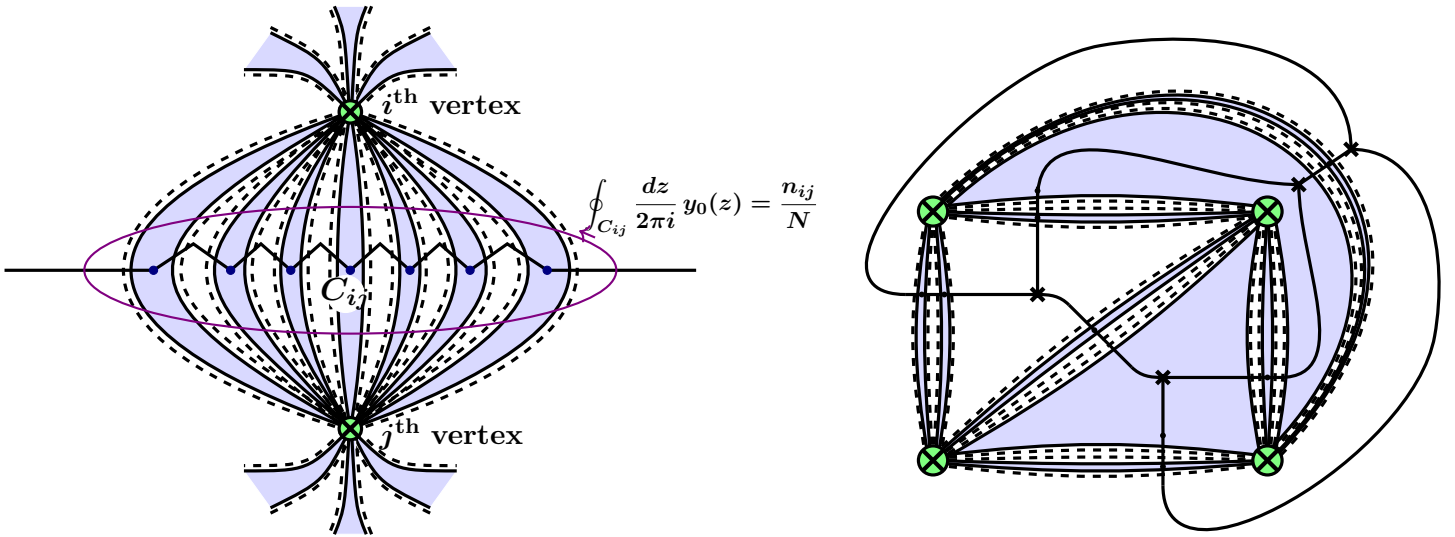

Rather remarkably, the quadratic differential formed from the spectral curve $- y_0^2(z) dz^2$
\begin{equation}
 - y_0^2(z) dz^2\equiv 4\pi^2 \phi_S(z)dz^2  
\end{equation}
defines a Strebel differential $\phi_S(z)dz^2$, a quadratic differential holomorphic everywhere except with double poles at $n$ marked points $z_i$ such that all the ``lengths" between its zeroes $\{ a_k \}$ are real 
\begin{equation}\label{streb-length}
l_{km}=\int_{a_k}^{a_m}\sqrt{\phi_S(z)} \in {\mathbb R}_+ \ .
\end{equation}
The latter condition is clearly satisfied due to Eq. (\ref{periodintegral}). Each Strebel differential defines a Strebel graph whose vertices are the zeroes of the differential and edges are trajectories connecting these zeroes such that the length of these trajectories are real (See fig. \ref{fig:triptych}). From our construction, we identify the Strebel graph corresponding to a covering map to the dual of the Feynman graph of that covering with its edges as the cuts of the spectral curve.  The significance of this connection lies in the theorem of Strebel \cite{Strebel}, which states, for every Riemann surface $\Sigma_{g,n}$ with genus $g$ and $n$ marked points where $n>0$ and $2g + n>2$, and any $n$ specified positive numbers $(w_1,\ldots,w_n)$, three exists a unique Strebel differential.  Thus the Strebel lengths parametrise the moduli space ${\cal M}_{g,n}$. We can therefore associate to each Strebel graph with its lengths, and thus Feynman graphs with fixed number of Wick contractions, to a given point in the moduli space.

In other words, in the large twist limit (with specified twists) $(w_1,\ldots,w_n)$, corresponding to each covering map specified by the periods of the A and B-cycles (or lengths of the edges of the Strebel graph) there exists a point in the (decorated) moduli space of the 
$n$-punctured sphere $\mathcal{M}_{0,n}$. The sum over coverings defining the symmetric orbifold correlator in Eq. (\ref{correlator}) goes over to an integral covering this moduli space exactly once, see figure \ref{fig:flowchart}.
\begin{figure}[!htb]
\hspace{-4.5mm}
 \includegraphics[width=1.05
    \textwidth]{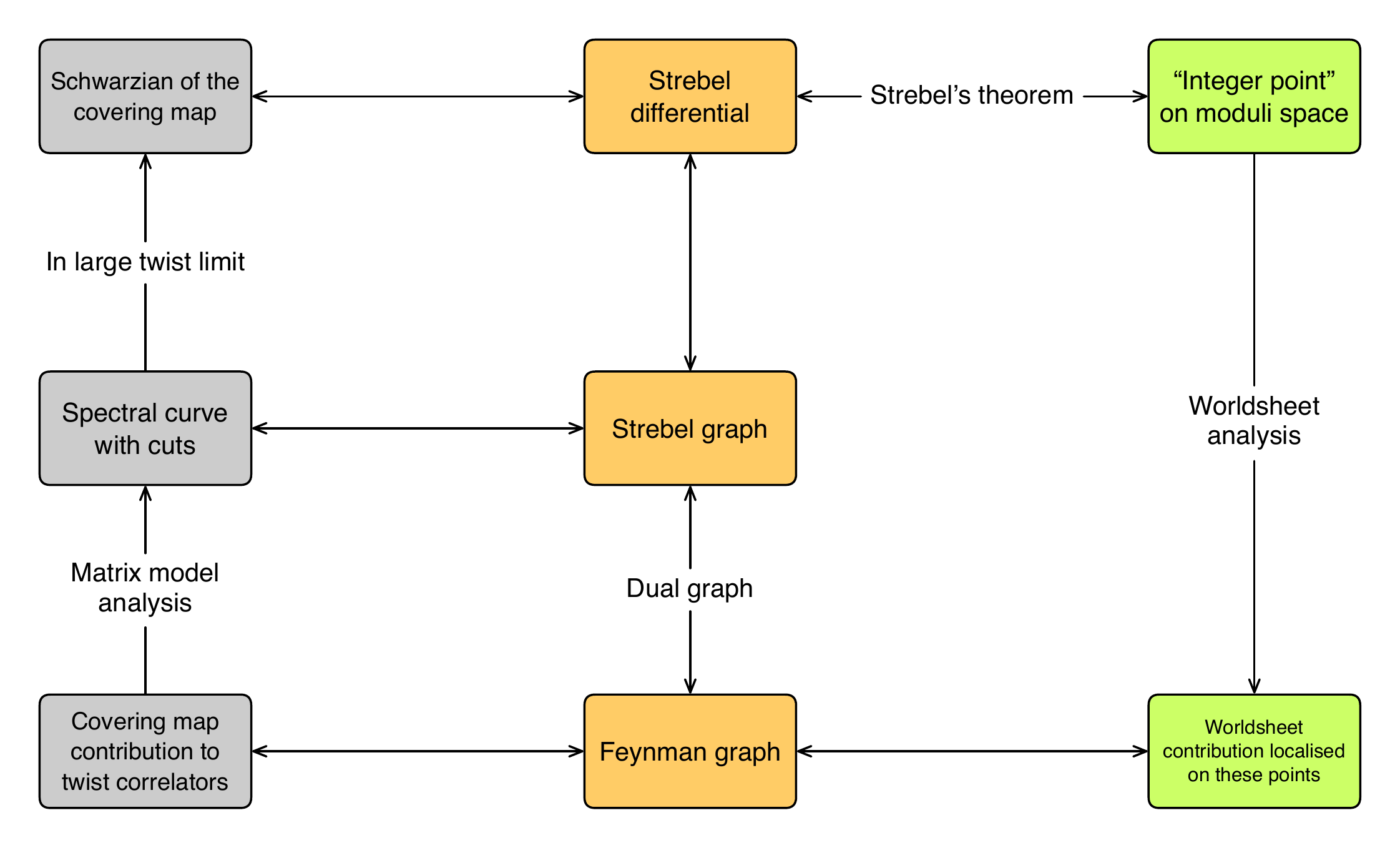}
\caption{Flowchart explaining the general logic of how a field theory correlator directly maps onto  a worldsheet integral over string moduli space where each Feynman diagram gets associated with a point on the moduli space. This is seen in the middle and right hand verticals of the figure. The left hand side indicates the specific way the equivalence is seen in the ${\rm AdS}_3$ case, where it proceeds via an auxiliary matrix model. This tool bridges the field theory and the string theory in this case.}
\label{fig:flowchart}
\end{figure}

\par 
The Lunin-Mathur weight of the Liouville action for the conformal factor also has a natural interpretation from the point of view of the worldsheet picture. One finds \cite{Gaberdiel:2020ycd}
\begin{equation}
S_L[\Gamma] = \frac{N^2}{8\pi} \int d^2z |\phi_S(z)| = \frac{1}{4\pi} \int d^2z \big{|} S[\Gamma]\big{|} \ .
\end{equation}
The first equality tells us that the action weighting each covering map is given by the Strebel area i.e. 
the worldsheet area when we fix the worldsheet metric to be in Strebel gauge  $ds^2=|\phi_S(z)|dz d\bar{z}$ \cite{Gopakumar:2011ev}.  
In the second equality, we have used the identification of the Strebel differential and the Schwarzian, \be\label{strebschw}
4\pi^2 \phi_S(z) = \frac{2}{ N^2} S[\Gamma] \ ,
\ee  
which holds in the strict large $N$ limit.
As we saw in Sec. \ref{sec.5}, such an identification holds in the case of ${\rm AdS}_5$ too, in the special kinematic configuration considered there.

\end{document}